\documentclass[%
 reprint,
superscriptaddress,
 amsmath,amssymb,
 aps,
]{revtex4-1}

\usepackage{mathrsfs}
\usepackage{amsmath}
\usepackage{braket}
\usepackage{graphicx}
\usepackage{dcolumn}
\usepackage{bm}
\usepackage{comment}
\usepackage{amsthm}
\theoremstyle{definition}

\newtheorem*{theorem*}{Prop}

\begin{document}

\preprint{APS/123-QED}

\title{Exact Floquet quantum many-body scars under Rydberg blockade}

\author{Kaoru Mizuta}
 \email{mizuta.kaoru.65u@st.kyoto-u.ac.jp}
\affiliation{%
 Department of Physics, Kyoto University, Kyoto 606-8502, Japan
}%
\author{Kazuaki Takasan}%
\affiliation{%
 Department of Physics, University of California, Berkeley, California 94720, USA
}%
\author{Norio Kawakami}
\affiliation{%
 Department of Physics, Kyoto University, Kyoto 606-8502, Japan
}%

\date{\today}
             

\begin{abstract}
Quantum many-body scars have attracted much interest as a violation of the eigenstate thermalization hypothesis (ETH) due to recent experimental observation in Rydberg atoms and related theoretical studies. In this paper, we construct a model hosting exact Floquet quantum many-body scars, which violate 
the Floquet version of ETH. We consider two uniformly driven static Hamiltonians prohibiting neighboring up spins (Rydberg blockade) like the PXP model, and construct a binary drive composed of them. We show that there exists a four-dimensional subspace which completely avoids thermalization to infinite temperature and that any other states, including some special scar states reported in the static PXP model, are vulnerable to heating and relax to infinite temperature. We also construct a more generalized periodic drive composed of time-dependent PXP-type Hamiltonians showing exact Floquet quantum many-body scars and discuss possible experimental realization of the model in Rydberg atoms.
\end{abstract}

\pacs{Valid PACS appear here}
\maketitle


\section{Introduction}
Thermalization in closed quantum systems has been vigorously studied to understand the relationship between quantum physics and statistical physics. With the recent progress in numerical and experimental studies \cite{Deutsch1991,Srednicki1994,Rigol2008,Bernien2017}, generic nonintegrable systems have been believed to satisfy the so-called eigenstate thermalization hypothesis (ETH). ETH dictates that all the eigenstates cannot be distinguished from thermal equilibrium states as long as only macroscopic observables are considered. Since ETH is a sufficient condition for thermalization to take place, ETH is believed to be a key to understand thermalization.

However, with the development of Rydberg atoms \cite{Bernien2017}, it has been revealed that there exists a violation of ETH called \textit{quantum many-body scars} \cite{Shiraishi2017}.  To be precise, there are several nonthermal eigenstates, which are eigenstates distinguishable from thermal equilibrium states (called exact scar eigenstates), while other states out of their subspace experience thermalization as with usual nonintegrable models. The PXP model, a typical model showing scars, has been realized on Rydberg atoms where adjacent atoms in Rydberg states are prohibited (Rydberg blockade) \cite{Turner2018,Turner2018B,Ho2019,Shiraishi2019}.  A nonthermalizing oscillation of domain-wall density,  which seems to be related to exact scar eigenstates \cite{Lin2019}, has been observed. A number of recent studies have found the existence of exact scar eigenstates also in other spin models \cite{Moudgalya2018,Moudgalya2018B,Schecter2019,Iadecola2020,Mark2020,Moudgalya2020,Shibata2020}.

In contrast, in periodically driven (Floquet) cases, nonintegrable systems are believed to satisfy the Floquet version of ETH (Floquet-ETH), which says that all the eigenstates of the time evolution operator for one period cannot be distinguished from a trivial infinite temperature state \cite{DAlessio2014,Lazarides2014}. This is a sufficient condition for any initial state to be thermalized to infinite temperature, which can be interpreted as a consequence from the absence of energy conservation. While thermal equilibrium states in static many-body systems cause various attractive phenomena such as spontaneous symmetry breaking, interacting Floquet systems often become trivial due to Floquet-ETH in the thermodynamic limit, except for a few examples such as Floquet many-body localization \cite{Po2016,Abanin2017} and Floquet time crystals \cite{Else2016,Keyserlingk2016,Khemani2016,Khemani2017,Yao2017,Choi2017,Zhang2017}. Thus, quantum many-body scars in Floquet systems (Floquet quantum many-body scars) are also of great interest as a violation of Floquet-ETH.

Some recent studies have tackled the realization of quantum many-body scars in Floquet systems \cite{Mukherjee2020,Pai2019,Khemani2019axv,Haldar2019,Sugiura2019}. The former references \cite{Pai2019,Khemani2019axv} consider a system dominated by random unitary matrices preserving charges and dipole moments,  and numerically \cite{Pai2019} and analytically \cite{Khemani2019axv} show the existence of states immune to thermalization. The latter one \cite{Sugiura2019} rigorously constructs Floquet quantum many-body scars realized by quasienergy-degeneracy modulo $2\pi$ (Floquet-intrinsic scars). 

In this paper, we demonstrate a systematic construction of exact Floquet quantum many-body scars. We consider a binary drive composed of uniformly driven PXP-type static models, and obtain a four-dimensional subspace which rigorously avoids thermalization to infinite temperature. In our construction, the instantaneous Hamiltonians share a subspace immune to thermalization, though they have different nonthermal scar eigenstates. Using these properties, we can realize exact Floquet quantum many-body scars showing persistent dynamics both stroboscopically and microscopically without fine-tuning of the switching time. We also show that these properties enable us to construct a generic periodic drive hosting exact Floquet quantum many-body scars. These results will shed light on understanding of nonequilibrium dynamics in Floquet systems.

\section{Models and Outline}
First, we identify the protocol of driving and show the outline of this paper. Floquet systems are described by a time-periodic Hamiltonian $H(t)$ which satisfies $H(t)=H(t+T)$ ($T$: period). Here, we focus on a binary drive as the simplest protocol, described by
\begin{equation}\label{BinaryDrive}
H(t) = \begin{cases}
H_1 & 0 \leq t < T_1 \\
H_2 & T_1 \leq t < T_1 + T_2 = T,
\end{cases}
\end{equation}
and then the Floquet operator $U_f$ (the time evolution operator for one period $T$) is given by
\begin{equation}
U_f = e^{-iH_2T_2} e^{-iH_1T_1}.
\end{equation}
What distinguishes Floquet systems from static systems is the noncommutative property of Hamiltonians at different time, and hence we assume
\begin{equation}
[H_1,H_2] \neq 0.
\end{equation}
To construct Floquet quantum many-body scars, we make use of two different static Hamiltonians $H_1$ and $H_2$ hosting quantum many-body scars. With choosing proper uniform PXP-type Hamiltonians, defined on a Hilbert space prohibiting adjacent up spins (Rydberg blockade), we rigorously show the existence of a four-dimensional subspace immune to relaxation to infinite temperature. 

Following the above strategy, this paper is organized as follows. In Section \ref{StaticModels}, we introduce two static models hosting quantum many-body scars. One of them is a well-known model called the PXP model. We construct the PY$_4$P model as another model inequivalent to the PXP model. In Section \ref{FloquetScar}, we construct a binary drive which shows Floquet quantum many-body scars using these static models, and numerically examine its nonintegrability to demonstrate that it can be a nontrivial example for the violation of Floquet-ETH. Then, we rigorously show the existence of an embedded subspace, which is a subspace completely immune to thermalization. We also demonstrate the real-time dynamics and show thermalization dependent on whether the initial states belong to the embedded subspace (Section \ref{FloquetThermal}). Finally, we discuss generalization of our binary drive to a generic time-periodic drive (Section \ref{Generalization}) and end up with discussing a possible experimental realization in Rydberg atoms and concluding this paper (Section \ref{DiscussionAndConclusion}).

\section{Static PXP and PY$_4$P models}\label{StaticModels}
In this section, we introduce the PXP model as a typical static model showing scars, and construct another model called the PY$_4$P model with its analogy. Throughout the paper, we consider a one-dimensional Ising chain under open boundary conditions (OBC). Assume that the number of the sites $L$ is a multiple of $4$. We consider a Hilbert space prohibiting states which include neighboring up spins, and then the dimension of the Hilbert space $\mathcal{D}_L$ is given by $\mathcal{D}_L=F_{L+2}$ ($F_n$ : the Fibonacci sequence) \cite{Sinfinite}. The PXP model under OBC is described by the following Hamiltonian \cite{Bernien2017,Turner2018}:
\begin{eqnarray}
H_X &=& \sum_{i=2}^{L-1} P_{i-1} X_i P_{i+1} + X_1 P_2 + P_{L-1} X_L, \label{H_X} \\
P_i &=& (1-Z_i)/2.
\end{eqnarray}
Here, we denote Pauli operators on the $i$-th site by $I_i$,$X_i$,$Y_i$ and $Z_i$. $P_i$ represents the projection to a down spin state on the $i$-th site. Such a Hamiltonian on the constrained Hilbert space is realizable in Rydberg atoms by quite strong repulsive interactions between adjacent atoms in Rydberg states \cite{Bernien2017}. The PXP model is known to violate ETH since the following four eigenstates $\ket{\Gamma_{\alpha\beta}^x}$ ($\alpha,\beta=1,2$)  are not thermal:
\begin{eqnarray}\label{mps_x}
\ket{\Gamma_{\alpha\beta}^x} &=& \sum_{\vec{\sigma}} \vec{u}_{\alpha}^\dagger B^{\sigma_1} C^{\sigma_2} \hdots B^{\sigma_{L-1}} C^{\sigma_L} \vec{u}_{\beta} \ket{\vec{\sigma}} \\
B^{\uparrow} &=& \sqrt{2} \left( \begin{array}{ccc}
0 & 0 & 0 \\
1 & 0 & 1 \end{array} \right), \quad
B^{\downarrow} = \left( \begin{array}{ccc}
1 & 0 & 0 \\
0 & 1 & 0 \end{array} \right), \\
C^{\uparrow} &=& \sqrt{2} \left( \begin{array}{ccc}
1 & 0 \\
0 & 0 \\
-1 & 0 \end{array} \right), \quad
C^{\downarrow} = \left( \begin{array}{cc}
0 & -1 \\
1 &  0 \\
0 & 0 \end{array} \right), \\
\vec{u}_1 &=& \frac{1}{\sqrt{2}} \left( \begin{array}{c}
1 \\
1 \end{array} \right), \qquad \quad
\vec{u}_2 = \frac{1}{\sqrt{2}} \left( \begin{array}{c}
1 \\
-1 \end{array} \right). \label{vector_x}
\end{eqnarray}
Here, the summation is taken over all possible spin configurations $\vec{\sigma}=(\sigma_1,\hdots,\sigma_L)$ with each spin $\sigma_i$ taking $\uparrow$ or $\downarrow$. These states do not include neighboring up spins due to $B^{\uparrow} C^{\uparrow}=O$ and $C^{\uparrow} B^{\uparrow}=O$, and they have the following eigenvalues,
\begin{eqnarray}\label{eigen_x}
\begin{aligned}
& H_X \ket{\Gamma_{11}^x} = 0, \quad
H_X \ket{\Gamma_{12}^x} = \sqrt{2} \ket{\Gamma_{12}^x}, \\
& H_X \ket{\Gamma_{22}^x} = 0, \quad
H_X \ket{\Gamma_{21}^x} = -\sqrt{2} \ket{\Gamma_{21}^x},
\end{aligned}
\end{eqnarray}
which are derived in Ref. \cite{Lin2019}.


Next, as a static model showing scars which is inequivalent to the PXP model, we construct the PY$_4$P model defined under the constrained Hilbert space, described by
\begin{eqnarray}
H_Y &=& \sum_{i=2}^{L-1} c_i  P_{i-1} Y_i P_{i+1} + Y_1 P_2 + P_{L-1} Y_L , \label{H_Y} \\
c_i &=& \sqrt{2} \cos \left( \frac{i\pi}{2} - \frac{\pi}{4} \right). \label{c_i}
\end{eqnarray}
This model possesses quadruple lattice-periodicity. We define four states $\ket{\Gamma_{\alpha\beta}^y}$ by
\begin{eqnarray}
\ket{\Gamma_{\alpha\beta}^y} &=& \sum_{\vec{\sigma}} \vec{v}_{\alpha}^\dagger B^{\sigma_1} C^{\sigma_2} \hdots B^{\sigma_{L-1}} C^{\sigma_{L}} \vec{v}_{\beta} \ket{\vec{\sigma}} \label{mps_y} \\
\vec{v}_1 &=& \frac{1}{\sqrt{2}} \left( \begin{array}{c}
1 \\
i \end{array} \right), \qquad
\vec{v}_2 = \frac{1}{\sqrt{2}} \left( \begin{array}{c}
1 \\
-i \end{array} \right). \label{vector_y}
\end{eqnarray}
The action of $H_Y$ on these four states can be calculated by using the properties of the matrices $B^\sigma$ and $C^\sigma$. In a similar way to the PXP model, we obtain the following relation:
\begin{eqnarray}
H_Y \ket{\Gamma_{\alpha\beta}^y} 
&=& \sum_{\vec{\sigma}} \vec{v}_\alpha^\dagger Y B^{\sigma_1} C^{\sigma_2} \hdots B^{\sigma_{L-1}} C^{\sigma_{L}}  \vec{v}_\beta \ket{\vec{\sigma}} \nonumber \\
&-&  \sum_{\vec{\sigma}} \vec{v}_\alpha^\dagger B^{\sigma_1} C^{\sigma_2} \hdots B^{\sigma_{L-1}} C^{\sigma_{L}}  Y \vec{v}_\beta \ket{\vec{\sigma}}, \label{YA_AY}
\end{eqnarray}
where the $2 \times 2$ matrix $Y$ is given by $Y=\sigma^y/\sqrt{2}$ ($\sigma^x,\sigma^y,\sigma^Z$: the Pauli matrices). The detailed derivation is provided in Appendix \ref{Appendix_PYP}. Using the fact that $\vec{v}_\alpha$ is the eigenvector of $Y$ with the eigenvalue $(-1)^{\alpha-1}/\sqrt{2}$, we obtain that the four states $\ket{\Gamma_{\alpha\beta}^y}$ are static exact scar eigenstates of the PY$_4$P model $H_Y$:
\begin{eqnarray}\label{eigen_y}
\begin{aligned}
& H_Y \ket{\Gamma_{11}^y} = 0, \quad
H_Y \ket{\Gamma_{12}^y} = \sqrt{2} \ket{\Gamma_{12}^y},  \\
& H_Y \ket{\Gamma_{22}^y} = 0, \quad
H_Y \ket{\Gamma_{21}^y} = -\sqrt{2} \ket{\Gamma_{21}^y}.
\end{aligned}
\end{eqnarray}

The PY$_4$P Hamiltonian $H_Y$ is related to the PXP Hamiltonian $H_X$ by a unitary transformation as follows:
\begin{equation}\label{UnitaryEquiv}
H_Y = \mathcal{U}_Z H_X \mathcal{U}_Z^\dagger, \quad
H_X = - \mathcal{U}_Z H_Y \mathcal{U}_Z^\dagger,
\end{equation}
\begin{equation}
\mathcal{U}_Z = \exp \left( -i \frac{\pi}{4} \sum_i c_i Z_i \right).
\end{equation}
Thus, properties of the PXP model as a static scar are inherited to the PY$_4$P model, including ithe nonintegrability, the violation of ETH, and the anomalously long nonthermalizing oscillation from a $\mathbb{Z}_2$-ordered state.

\begin{figure*}
\hspace{-1cm}
\begin{center}
    \includegraphics[height=5.5cm, width=18cm]{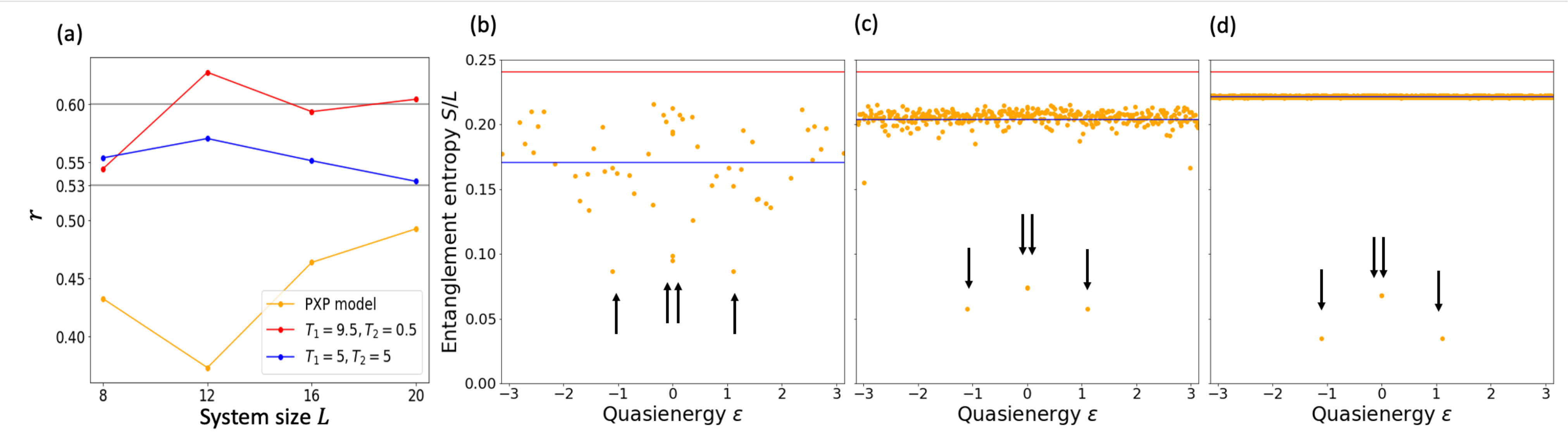}
    \caption{(a) Level statistics of the PXP model (the yellow line) and the periodically driven model (the red and blue lines). In the Floquet case, it rapidly approaches a value close to $0.6$ ($T_1=9.5, T_2=0.5$) or $0.53$ ($T_1=T_2=5$, with time-reversal symmetry) as the system size grows. Both results imply the nonintegrability of the driven model. (b)-(d): Entanglement entropy per length for each Floquet eigenstate of a different system size (b) $L=8$, (c) $L=12$, and (d) $L=20$. We use $T_1=9.5$ and $T_2=0.5$ as the parameters. The blue (lower solid) lines, representing the mean values of entanglement entropy, approach the red (upper solid) lines which denote the one at infinite temperature as the system size increases. The four marked states designated by the arrows (two points are degenerated at $\varepsilon=0$) remain low-entangled since they are exact scar eigenstates of the periodically driven model within the embedded subspace $\mathcal{S}$.} 
    \label{Fig_level_entanglement}
 \end{center}
 \end{figure*}

\section{Floquet Quantum Many-Body Scars}\label{FloquetScar}
In this section, we construct a binary drive showing Floquet quantum many-body scars. To confirm that our model becomes a nontrivial example violating Floquet-ETH, we numerically demonstrate nonintegrability of our model, and rigorously prove the existence of exact Floquet scar eigenstates, which are distinguishable from infinite temperature states.

\subsection{Model}\label{Model}
Assume that the system is a one-dimensional Ising chain where pairs of neighboring up spins are prohibited. Then, we consider a binary drive composed of the static Hamiltonians in the previous section:
\begin{equation}\label{BinaryXY}
H(t) = \begin{cases}
H_X & 0 \leq t < T_1 \\
H_Y & T_1 \leq t < T_1 + T_2 = T,
\end{cases}
\end{equation}
and then its Floquet operator is written by
\begin{equation}\label{FloquetOpXY}
U_f = e^{-iH_Y T_2} e^{-iH_XT_1}.
\end{equation}
Here, for Floquet quantum many-body scars to take place, $T_1$ and $T_2$ are arbitrary except for the case when either one of them is zero (there is no need for fine-tuning of them). 

We note the symmetries underlying this model. First, it possesses an inversion symmetry $I$ which maps each $i$-th site to the $(L-i+1)$-th site, and the Floquet operator $U_f$ is invariant under the inversion. Second, a  nonlocal chiral symmetry $\mathcal{C}$, designated by
\begin{equation}\label{ChiralSymmetry}
\mathcal{C} U_f \mathcal{C}^\dagger = U_f^\dagger, \quad \mathcal{C} = \left( \prod_i Z_i \right) e^{iH_YT_2},
\end{equation}
is also respected. This chiral symmetry makes the spectrum of quasienergy $\{ \varepsilon \}$ (the eigenvalues of $-i\log U_f$) symmetric to $\varepsilon=0$. In addition, if $T_1=T_2$ is satisfied, the model also respects a time-reversal symmetry (TRS), described by
\begin{equation}\label{TimeReversalSymmetry}
\mathcal{U}_Z U_f^\ast \mathcal{U}_Z^\dagger = U_f^\dagger, \quad \mathcal{U}_Z = \exp \left( - i \frac{\pi}{4} \sum_i c_i Z_i \right),
\end{equation}
where we use the relation Eq. (\ref{UnitaryEquiv}).

\subsection{Nonintegrability}\label{Nonintegrablity}
To confirm that the model can be a nontrivial example of violation of Floquet-ETH, we begin with analyzing nonintegrability of the model. Considering the nonintegrability of the PXP model, that of  the PY$_4$P model \cite{Turner2018,Khemani2019}, and the noncommutability $[H_X,H_Y]\neq0$, the model is also expected to be nonintegrable. Here, we demonstrate this by calculating level statistics. Using the $n$-th quasienergy $\varepsilon_n$ (replaced by eigenenergy $E_n$ in static cases) and its gap $\Delta_{n}=\varepsilon_{n+1}-\varepsilon_n$, let us define the level spacing ratio $r_n$ by $r_n=\min (\Delta_n/\Delta_{n+1},\Delta_{n+1}/\Delta_{n})$ and denote their spectrally averaged value by $r \equiv \left<r_n \right>$. When the model is nonintegrable with the increasing system size,  $r$ approaches a Gaussian orthogonal ensemble value close to $0.53$ if it is time-reversal symmetric, or approaches a Gaussian unitary ensemble value close to $0.6$ otherwise \cite{Oganesyan2007,Atas2013,Khemani2019}. On the other hand, in integrable systems, it approaches a value close to $0.39$, that of Poisson statistics. Figure \ref{Fig_level_entanglement} (a) shows the numerical result for $r$ calculated by exact diagonalization (ED). Considering the inversion symmetry $I$, we limit the Hilbert space to the inversion-plus sector, the subspace the eigenvalue of $I$ of which is $+1$. The red upper solid line (blue middle solid line) represents the case of $T_1=9.5, T_2=0.5$ without TRS (that of $T_1=T_2=5$ with TRS) respectively. In each case, $r$ flows to a value close to $0.6$ (a value close to $0.53$) as the system size grows, and hence we can conclude the nonintegrability of the model.

We also demonstrate the nonintegrability in terms of entanglement entropy of each Floquet eigenstate. Entanglement entropy of a given state $\ket{\psi}$ is defined by
\begin{equation}
S[\psi]=-\mathrm{Tr}_A [\rho_A \log \rho_A], \quad \rho_A = \mathrm{Tr}_B \ket{\psi}\bra{\psi},
\end{equation}
where the subsystem $A$ ($B$) represents the left (right) half of the system.  A state $\ket{\psi}$ indistinguishable from the infinite temperature state is expected to possess entanglement entropy equal to that of infinite temperature $S_\infty$, and hence it obeys a volume law in the thermodynamic limit as follows \cite{Sinfinite}:
\begin{equation}
\lim_{L\to\infty}\frac{S_\infty}{L} =\frac{1}{2} \log \phi, \quad \phi=\frac{1+\sqrt{5}}{2}.
\end{equation}
Figure \ref{Fig_level_entanglement} (b)-(d) shows the numerical results at $T_1=9.5$ and $T_2=0.5$. 
The red lines are entanglement entropy per volume at infinite temperature, while the blue lines represent the averaged entanglement entropy per volume of all the  Floquet eigenstates. As the system size $L$ increases, Floquet eigenstates become featureless, with its entanglement entropy approaching the one at infinite temperature. On the other hand, there exist four anomalous low-entangled Floquet eigenstates designated by the four marked points in the figure. As discussed in the following section, these four states are nothing but exact Floquet scar eigenstates, which are distinguishable from the infinite temperature state. This result elucidates both the nonintegrability and the existence of nontrivial Floquet quantum many-body scars.


\begin{figure*}
\hspace{-1cm}
\begin{center}
    \includegraphics[height=6.5cm, width=18cm]{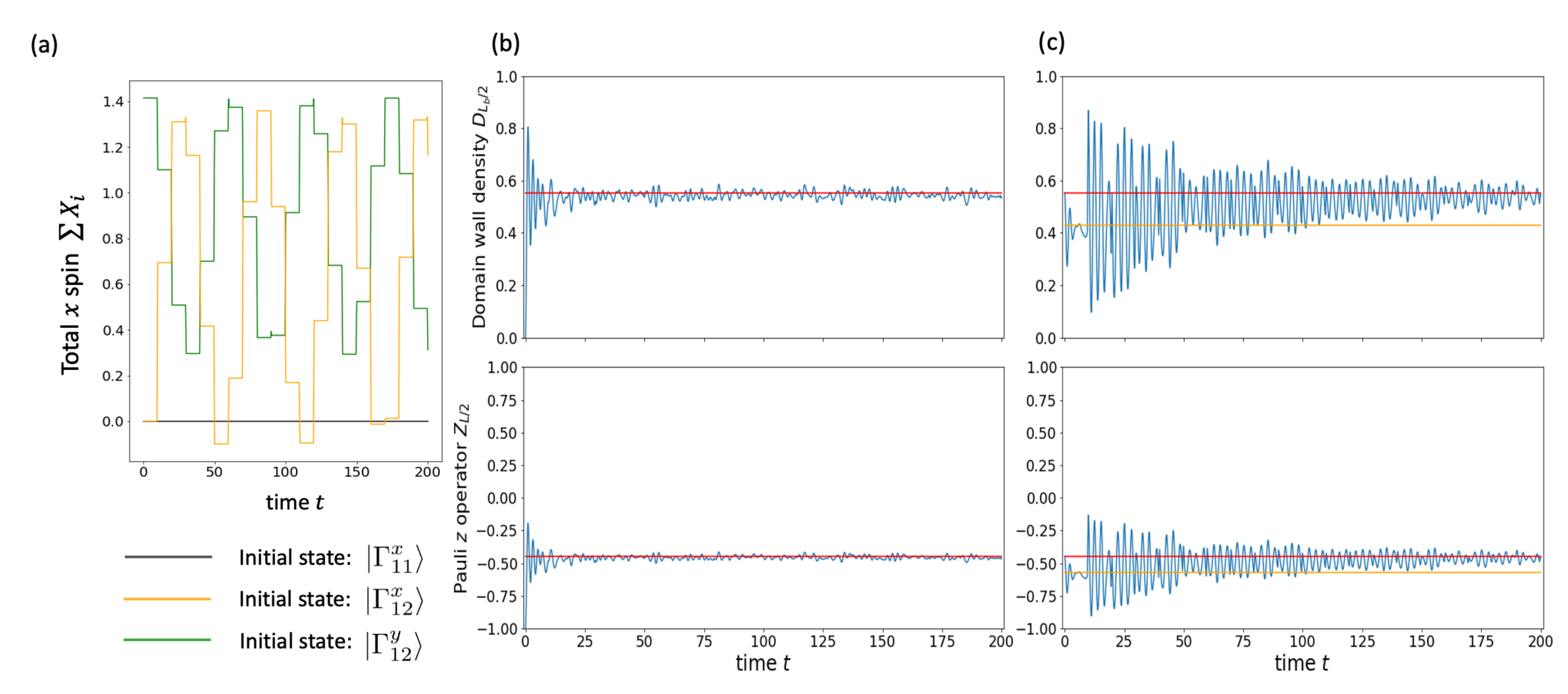}
    \caption{(a) Real-time dynamics of total $x$ spin $\sum_i X_i$ under the initial states within $\mathcal{S}$ at $T_1=9.5, T_2=0.5$. The states $\ket{\Gamma_{12}^x}$ and $\ket{\Gamma_{12}^y}$ show a persistent oscillation. (b) Real-time dynamics under the initial state $\ket{\psi_1}$, which is at infinite temperature under $H_X$. The red lines represent the values at infinite temperature. (c) Real-time dynamics under the initial state $\ket{\psi_2}$, which is at finite temperature $\beta_\text{eff}$ under $H_X$. The lower solid yellow lines represent the finite-temperature-equilibrium values under $H_X$ [$\beta_{\text{eff}}$ is obtained by numerically solving Eq. (\ref{beta_eq})]. The Floquet drive breaks such a feature of the initial state, and makes the observable approach the values at infinite temperature (the upper solid red lines). } 
    \label{Fig_dynamics}
 \end{center}
 \end{figure*}

\subsection{Exact Floquet scar eigenstates}\label{Sec_ExactScar}
We rigorously show a violation of Floquet-ETH in our model. To be precise, we prove that there exist four exact Floquet scar eigenstates, which can be distinguished from infinite temperature states. We define $\mathcal{S}$ by the four-dimensional subspace spanned by $\{ \ket{\Gamma_{\alpha\beta}^x} \}_{\alpha,\beta=1,2}$. Within the subspace $\mathcal{S}$, thermalization does not take place under $\mathrm{exp} (-iH_X T_1)$ by its definition. On the other hand, using Eqs. (\ref{mps_x}) and (\ref{mps_y}), we obtain
\begin{eqnarray}\label{eigen_yx}
\begin{aligned}
\ket{\Gamma_{11}^y} &=& \frac{1}{2} ( \ket{\Gamma_{11}^x} +i \ket{\Gamma_{12}^x} - i \ket{\Gamma_{21}^x} + \ket{\Gamma_{22}^x} ), \\
\ket{\Gamma_{12}^y} &=& \frac{1}{2} ( i \ket{\Gamma_{11}^x} + \ket{\Gamma_{12}^x} +\ket{\Gamma_{21}^x} - i \ket{\Gamma_{22}^x} ), \\
\ket{\Gamma_{21}^y} &=& \frac{1}{2} ( -i\ket{\Gamma_{11}^x} + \ket{\Gamma_{12}^x} + \ket{\Gamma_{21}^x} + i \ket{\Gamma_{22}^x} ), \\
\ket{\Gamma_{22}^y} &=& \frac{1}{2} ( \ket{\Gamma_{11}^x} -i \ket{\Gamma_{12}^x} + i \ket{\Gamma_{21}^x} +\ket{\Gamma_{22}^x} ).
\end{aligned}
\end{eqnarray}
Since this transformation is invertible, the subspace $\mathcal{S}$ is identical to the one spanned by $\{ \ket{\Gamma_{\alpha\beta}^y} \}_{\alpha,\beta=1,2}$.  Thus, thermalization does not take place in the subspace $\mathcal{S}$ also under $\exp(-i H_Y T_2)$, and hence we can conclude the absence of thermalization in the subspace $\mathcal{S}$ under the Floquet operator $U_f$. We note that this behavior can be understood also from the existence of local conserved quantities in the embedded subspace $\mathcal{S}$ as discussed in Appendix \ref{Appendix_LocalConserved}. 

The existence of the four-dimensional embedded subspace ensures the existence of four exact Floquet scar eigenstates. Two of them given by
\begin{eqnarray}
\ket{\Gamma_0} &=& \ket{\Gamma_{11}^x}+\ket{\Gamma_{22}^x}, \label{Gamma_zero} \\
\ket{\Gamma_0^\prime} &=& \sin\frac{T_1}{\sqrt{2}} \cos\frac{T_2}{\sqrt{2}} (\ket{\Gamma_{11}^x}-\ket{\Gamma_{22}^x}) \nonumber \\
&\quad& + i \sin \frac{T_2}{\sqrt{2}} (e^{iT_1/\sqrt{2}}\ket{\Gamma_{12}^x}-e^{-iT_1/\sqrt{2}}\ket{\Gamma_{21}^x}), \label{tilde_Gamma_zero}
\end{eqnarray}
have quasienergy zero and are invariant under the chiral symmetry operation $\mathcal{C}$ modulo constant. We can also obtain the analytical forms of the other two eigenstates, but we do not show them since their explicit forms are too complicated to write down and would not be useful. Instead, we find that the two eigenstates are related to each other by $C$ and hence they have quasienergies with the opposite signs. Moreover, since $C$ becomes an onsite symmetry within $\mathcal{S}$ as discussed in Appendix \ref{Appendix_LocalConserved}, they have the same entanglement entropy, while other pairs outside of $\mathcal{S}$ do not.

All the four exact Floquet scar eigenstates appear in Fig. \ref{Fig_level_entanglement} (b)-(d) as the four marked low-entangled states. As discussed above, the two of them, lying in $\varepsilon \neq 0$, appear symmetrically with respect to $\varepsilon = 0$. Since each of the Floquet scar eigenstates is a superposition of the four states $\ket{\Gamma_{\alpha\beta}^x}$ with bond dimension 2, they are represented by matrix product states with at-most bond dimension 8 using the direct sum of matrices. Thus, their entanglement entropy per length decays with $O(1/L)$, which implies the nonthermal behavior of them. This result also corresponds to the numerical result [See Fig. \ref{Fig_level_entanglement} (b)-(d)]. 

Finally, to confirm the violation of Floquet ETH, we examine whether Floquet eigenstates within the subspace $\mathcal{S}$ can be distinguished from the infinite temperature state. Here, we focus on a domain-wall density defined by
\begin{equation}
D_b \equiv \frac{1}{2} (I-Z_{2b-1}Z_{2b}).
\end{equation}
Then, the expectation value of $D_b$ is obtained as
\begin{equation}
\lim_{L_b \to \infty} \braket{ \psi | D_b | \psi} =\frac{2}{3}
\end{equation}
for any state $\ket{\psi} \in \mathcal{S}$, $\braket{\psi | \psi} =1$ including the renormalized eigenstates of $U_f$ within $\mathcal{S}$ \cite{Sinfinite}. On the other hand, the expectation value of domain-wall density at infinite temperature in the thermodynamic limit is $2/\sqrt{5}\phi=0.542\hdots \neq 2/3$ ($\phi$: the golden ratio) under OBC \cite{Sinfinite}. Therefore, it can be concluded that the model Eq. (\ref{BinaryXY}) violates Floquet-ETH.

\section{Real-time Dynamics}\label{FloquetThermal}

We discuss the real-time dynamics in and out of the embedded subspace $\mathcal{S}$, which possibly leads to experimental detection of Floquet quantum many-body scars.

First, let us consider the real-time dynamics within the embedded subspace $\mathcal{S}$. We define renormalized scar eigenstates by $\ket{\tilde{\Gamma_{\alpha\beta}^x}} \equiv \ket{\Gamma_{\alpha\beta}^x}/|| \ket{\Gamma_{\alpha\beta}^x} ||$, and then $\{ \ket{\tilde{\Gamma_{11}^x}}, \ket{\tilde{\Gamma_{12}^x}}, \ket{\tilde{\Gamma_{21}^x}}, \ket{\tilde{\Gamma_{22}^x}} \}$ composes an orthonormal basis of $\mathcal{S}$ in the thermodynamic limit while $\ket{\Gamma_{11}^x}$ is not orthogonal to $\ket{\Gamma_{22}^x}$ \cite{Sembedded}. With this basis, the Floquet operator $U_f |_\mathcal{S}$ is represented by
\begin{equation}
U_f |_\mathcal{S} = \left( \begin{array}{cccc}
p & qr & qr^\ast & 1-p \\
-q & pr & -(1-p)r^\ast & q \\
-q & -(1-p)r & pr^\ast & q \\
1-p  & -qr & -qr^\ast & p
\end{array}\right),
\end{equation}
\begin{equation}
p = \frac{1+\cos\sqrt{2}T_2}{2}, \quad q=\frac{\sin\sqrt{2}T_2}{2}, \quad r=e^{-i\sqrt{2}T_1},
\end{equation}
and stroboscopic dynamics of any observable is determined by its matrix representation. For example, the total magnetization in the $x$ direction, $\lim_{L\to\infty} \sum_i X_i$, is given by $\mathrm{diag} (0,\sqrt{2},-\sqrt{2},0)$ and shows a persistent oscillation in general, while the local Pauli operator $Z_i$ and the domain-wall density $D_b$ remain constant since they are proportional to identity in $\mathcal{S}$ \cite{Sembedded}. On the other hand, concerning the microscopic dynamics, generic initial states in $\mathcal{S}$, different from $\ket{\Gamma_0}$, show some persistent motion since $\ket{\Gamma_0}$ is the unique simultaneous eigenstate of $H_X$ and $H_Y$ in $\mathcal{S}$. We show typical real-time dynamics in Fig. \ref{Fig_dynamics} (a).

Next, we demonstrate the behavior outside of the embedded subspace $\mathcal{S}$. Following the nonintegrability of the model, generic initial states are expected to relax to infinite temperature, and we numerically confirm it by ED [See Fig. \ref{Fig_dynamics} (b), (c)]. We consider two different initial states $\ket{\psi_1} \equiv \ket{\downarrow\downarrow\hdots}$ and $\ket{\psi_2} \equiv \mathcal{P}_K \ket{-}^{\otimes L}/\sqrt{\mathcal{D}_L}$, where we define $\ket{-}$ by $(\ket{\uparrow}-\ket{\downarrow})/\sqrt{2}$ and denote the projection to the constrained Hilbert space prohibiting adjacent up spins as $\mathcal{P}_K$.  They have an exponentially small overlap with $\mathcal{S}$ in terms of the system size, and $\ket{\psi_1}$ ($\ket{\psi_2}$) is an infinite-temperature state (a state with finite temperature $\beta_\text{eff}$) under the PXP Hamiltonian $H_X$ because of $\braket{\psi_1 | H_X | \psi_1} =0$ ($\braket{\psi_2 | H_X | \psi_2} =-2L/\sqrt{5}\phi$). Here, the temperature of a state $\ket{\psi}$ is determined by solving the energy conservation
\begin{equation} \label{beta_eq}
\frac{\braket{\psi | H_X | \psi}}{\braket{\psi | \psi}} = \frac{\mathrm{Tr} [ H_X e^{-\beta H_X} ]}{\mathrm{Tr} [ e^{-\beta H_X} ]}
\end{equation}
in terms of $\beta$. Figure \ref{Fig_dynamics} (b) and (c) show the dynamics at $T_1=9.5$ and $T_2=0.5$, which we choose so that pre-equilibration under an effective static Hamiltonian in the high-frequency regime can be avoided \cite{Kuwahara2016,Mori2016,Abanin2017B,Abanin2017Mat}. The model shows thermalization to infinite temperature regardless of initial states outside of the embedded subspace in contrast to the static PXP model and the PY$_4$P model, where the system relaxes to thermal states with a certain temperature depending on its initial states.

In the static PXP model, there also exist some special states such as $\ket{\mathbb{Z}_2}\equiv \ket{\uparrow\downarrow\uparrow\downarrow\hdots}$ which show nonthermalizing behaviors of observables (e.g. domain-wall density), though their overlap with the embedded subspace is small enough \cite{Bernien2017,Turner2018,Lin2019,Ho2019}. We demonstrate the existence of such special states in the periodically driven model in Appendix \ref{SpecialScar}. The numerical result says that such a special state is thermalized to infinite temperature as well as other generic initial states out of the embedded subspace. In the static PXP model and the PY$_4$P model, three types of dynamics--- complete absence of thermalization within the embedded subspace, seemingly nonthermalizing behavior of some special states, and thermalization of other generic states are observed. By contrast, we conclude that the periodically driven model only shows complete absence of thermalization to infinite temperature within the embedded subspace or thermalization of other generic states.

\section{Generalization}\label{Generalization}
We generalize our binary drive to generic time-periodic drives. For this purpose, we first introduce another static model, the PZ$_4$P model defined by
\begin{equation}
H_Z = - \sqrt{2} \left( \sum_{i=2}^{L-1} c_i P_{i-1} Q_i P_{i+1} + Q_1 P_2 + P_{L-1} Q_L  \right), \label{H_Z}
\end{equation}
\begin{equation}
Q_i = I_i - P_i = (1+Z_i)/2.
\end{equation}
Then, the PZ$_4$P model  possesses the following eigenstates:
\begin{eqnarray}\label{eigen_z}
\begin{aligned}
& H_Z \ket{\Gamma_{11}^z} = 0, \quad
H_Z \ket{\Gamma_{12}^z} = \sqrt{2}  \ket{\Gamma_{12}^z}, \label{eigen_z1} \\
& H_Z \ket{\Gamma_{22}^z} = 0, \quad
H_Z \ket{\Gamma_{21}^z} = -\sqrt{2}  \ket{\Gamma_{21}^z},  \label{eigen_z2}
\end{aligned}
\end{eqnarray}
where $\{ \ket{\Gamma_{\alpha\beta}^z} \}$ is given by
\begin{eqnarray}
\ket{\Gamma_{\alpha\beta}^z} &=& \sum_{\vec{\sigma}} \vec{w}_{\alpha}^\dagger B^{\sigma_1} C^{\sigma_2} \hdots B^{\sigma_{L-1}} C^{\sigma_{L}}  \vec{w}_{\beta} \ket{\vec{\sigma}}, \label{mps_z} \\
\vec{w}_1 &=& \left( \begin{array}{c}
1 \\
0 \end{array} \right), \qquad
\vec{w}_2 = \left( \begin{array}{c}
0 \\
1 \end{array} \right). \label{vector_z}
\end{eqnarray}
This derivation is given in Appendix \ref{Appendix_PZP}. Since the Hamiltonian $H_Z$ commutes with $Z_i$ for every $i$, the PZ$_4$P model is integrable and does not host nontrivial phenomena by itself. However, by combining the PXP Hamiltonian and the PY$_4$P Hamiltonian, we can construct nontrivial models described by
\begin{equation}\label{H_a}
H_{\vec{a}}= \vec{a} \cdot \vec{H}, \quad \vec{H} = (H_X,H_Y,H_Z)
\end{equation}
with $\vec{a} = (\sin\theta \cos \varphi, \sin\theta \sin \varphi, \cos \theta)$. This generalized model is no longer unitarily equivalent to the PXP model or PY$_4$P model like Eq. (\ref{UnitaryEquiv}). We can compose four exact scar eigenstates of this Hamiltonian, given by
\begin{eqnarray}
\ket{\Gamma_{\alpha\beta}^{\vec{a}}} &=& \sum_{\vec{\sigma}} \vec{u}_{\alpha}^{a,\dagger} B^{\sigma_1} C^{\sigma_2} \hdots B^{\sigma_{L-1}} C^{\sigma_{L}} \vec{u}_{\beta}^a \ket{\vec{\sigma}},
\end{eqnarray}
\begin{equation}
\vec{u}^a_1 = \left( \begin{array}{c}
\cos (\theta/2) \\
e^{i\varphi} \sin (\theta/2) \end{array} \right), \quad
\vec{u}^a_2 = \left( \begin{array}{c}
-e^{-i\varphi}\sin (\theta/2) \\
\cos (\theta/2) \end{array} \right). \label{vector_a}
\end{equation}
The eigenvalue of $\ket{\Gamma_{\alpha\beta}^{\vec{a}}}$ is $\{ (-1)^{\alpha-1} - (-1)^{\beta-1} \}/\sqrt{2}$. Using Eq. (\ref{YA_AY}) and similar relations for the PXP model \cite{Lin2019,Shiraishi2019} and the PZ$_4$P model [Appendix \ref{Appendix_PZP}] results in
\begin{equation}
H_{\vec{a}} \ket{\Gamma_{\alpha\beta}^{\vec{a}}} =
\sum_{\vec{\sigma}} \vec{u}_{\alpha}^{a,\dagger}  \left\{ (\vec{a} \cdot \vec{\Sigma}) D^{\vec{\sigma}} -D^{\vec{\sigma}} (\vec{a} \cdot \vec{\Sigma}) \right\} \vec{u}_{\beta}^a \ket{\vec{\sigma}} ,
\end{equation}
in which we define $\vec{\Sigma}=(\sigma^x,\sigma^y,\sigma^z)/\sqrt{2}$ and $D^{\vec{\sigma}}=B^{\sigma_1}C^{\sigma_2}\hdots B^{\sigma_{L-1}} C^{\sigma_L}$. Then we can confirm that $\ket{\Gamma_{\alpha\beta}^{\vec{a}}}$ become exact scar eigenstates by the fact that the vectors $\vec{u}_\alpha^a$ are eigenvectors of $\vec{a} \cdot \vec{\Sigma}$ with eigenvalues $(-1)^{\alpha-1}/\sqrt{2}$.

Now, we are ready to construct a generalized version of the binary drive Eq. (\ref{BinaryXY}), represented by
\begin{equation}
H(t)=\vec{a}(t) \cdot \vec{H}, \quad \vec{a}(t+T)=\vec{a}(t).
\end{equation}
Since $\ket{\Gamma_{\alpha\beta}^{\vec{a}}}$ is represented by a linear combination of $\{ \ket{\Gamma_{\alpha\beta}^{x}} \}$ for arbitrary $\vec{a}$, the dynamics under $H(t)$ is closed within the subspace spanned by $\{ \ket{\Gamma_{\alpha\beta}^{x}} \}$. Thus, this model always has a four-dimensional embedded subspace $\mathcal{S}$. Using the fact that $\{ \vec{u}_\alpha^a \}_{\alpha=1,2}$ is a complete orthonormal basis of $\mathbb{C}^2$,
\begin{eqnarray}
\ket{\Gamma_0} &\equiv& \ket{\Gamma_{11}^{\vec{a}}} + \ket{\Gamma_{22}^{\vec{a}}} \nonumber \\
&=& \sum_{\vec{\sigma}} \mathrm{Tr} \left[ B^{\sigma_1} C^{\sigma_2} \hdots B^{\sigma_{L-1}} C^{\sigma_{L}} \right]\ket{\vec{\sigma}},
\end{eqnarray}
which is independent of the choice of $\vec{a}$, is an eigenstate of $H_{\vec{a}}$ the eigenvalue of which is zero. Thus, we always provide one of the four exact Floquet scar eigenstates of $H(t)$ with $\ket{\Gamma_0}$, the quasienergy of which is zero. 

\section{Discussion and Conclusions}\label{DiscussionAndConclusion}
Before concluding the paper, let us briefly discuss how to realize the Hamiltonian (\ref{BinaryXY}) showing Floquet quantum many-body scars. The PXP model, which hosts static quantum many-body scars, is experimentally realized in Rydberg atoms \cite{Bernien2017}. Each atom can occupy the ground state $\downarrow$ and the Rydberg state $\uparrow$, which is an excited state with a large quantum number. Since the repulsive interactions between neighboring atoms in the Rydberg state are quite large, neighboring $\uparrow\uparrow$ pairs are prohibited. The Rabi oscillation in this limited subspace results in the PXP Hamiltonian $H_X$.

Once the PXP Hamiltonian $H_X$ is realized, our model is also realizable. We consider a potential with quadruple periodicity of the lattice,
\begin{equation}
Z_4 = \sum_{i} c_i Z_i,
\end{equation}
where the coefficients $c_i$ are given by Eq. (\ref{c_i}). Then, using the unitary equivalence Eq. (\ref{UnitaryEquiv}), the Floquet operator $U_f$ is
\begin{eqnarray}
U_f &=& e^{-iH_YT_2} e^{-iH_XT_1} \nonumber \\
&=& e^{-i Z_4 (\pi/4)} e^{-iH_XT_2} e^{-i(-Z_4)(\pi/4)} e^{-iH_XT_1}.
\end{eqnarray}
We can realize the PZ$_4$P model in a similar setting since the potential $Q_i$ acting only on Rydberg states becomes equivalent to each term $P_{i-1}Q_iP_{i+1}$ in the constrained Hilbert space. This is confirmed by the relation
\begin{equation}
\mathcal{P}_K Q_i \mathcal{P}_K = \mathcal{P}_K P_{i-1}Q_iP_{i+1} \mathcal{P}_K,
\end{equation}
where $\mathcal{P}_K=\prod_i (1-Q_i Q_{i+1})$ represents the projection operator to the constrained Hilbert space. Thus, our periodically driven models, including the generalized versions, can be realized  in Rydberg atoms. While our model does not require the fine-tuning of the parameters, what seems to be difficult is to prepare the initial state in the embedded subspace $\mathcal{S}$. However, considering that the states in the embedded subspace $\mathcal{S}$ are equivalent to the Affleck-Kennedy-Lieb-Tasaki state \cite{Lin2019} and that the exact scar states can show slow thermalization compared to other states even under local perturbations \cite{Lin2019axv}, the observation of the Floquet quantum many-body scars would be physically feasible in the near future.

In summary, we have constructed a nonintegrable model which hosts Floquet quantum many-body scars, driven by uniformly imposed Hamiltonians on the constrained Hilbert space prohibiting adjacent pairs of up spins. We have rigorously shown that the model violates Floquet-ETH with the fact that instantaneous Hamiltonians share a subspace immune to thermalization although the scar eigenstates do not correspond to one another. The entanglement spectrum of Floquet eigenstates and the real-time dynamics of the model indicate that any initial state outside of the embedded subspace is thermalized to infinite temperature.  We have also discussed a possible experimental realization of the model in Rydberg atoms, and thereby our result would contribute to understanding how closed Floquet systems equilibrate to infinite temperature.

\begin{acknowledgments}
We would like to thank H. Katsura for constructive comments on the derivation of the exact scar eigenstates and noticing us some mistakes in our calculation. This work is supported by a Grant-in-Aid for Scientific
Research on Innovative Areas ``Topological Materials Science''
(KAKENHI Grant No. JP15H05855) and also Japan Society for the Promotion of Science (KAKENHI
Grants No. JP18H01140, JP19H01838, and JP20J12930). K. M. is supported by WISE Program, MEXT, and a Research Fellowship for Young Scientists from JSPS.
K. T. thanks JSPS for support from Overseas Research Fellowship.
\end{acknowledgments}


\begin{thebibliography}{43}
\bibitem [{Sin()}]{Sinfinite}%
  \BibitemOpen
  \href@noop {} {}\bibinfo {note} {See Section A2 in Appendix. We briefly
  review local observables and entanglement entropy of the infinite temperature
  state in the constrained Hilbert space.}\BibitemShut {Stop}%
\bibitem [{Sem()}]{Sembedded}%
  \BibitemOpen
  \href@noop {} {}\bibinfo {note} {See Section A1 in Appendix. This provides
  exact scar eigenstates of the PY$_4$P model and the PZ$_4$P model. We also
  describe properties of the embedded subspace $\mathcal{S}$.}\BibitemShut
  {Stop}%
\end{thebibliography}

\providecommand{\noopsort}[1]{}\providecommand{\singleletter}[1]{#1}%

\clearpage

\newcommand{\wsq}{\qquad $\square$}
\newtheorem*{proof*}{Proof}
\newtheorem{lemma}{Lemma}
\renewcommand{\thesection}{A\arabic{section}}
\renewcommand{\theequation}{A\arabic{equation}}
\setcounter{equation}{0}
\setcounter{section}{0}

\onecolumngrid
\begin{center}
 {\large 
 {\bfseries Appendix }}
\end{center}
\vspace{10pt}

\section{Properties of Embedded Subspace}

\subsection{Exact scar eigenstates of the PY$_4$P model}\label{Appendix_PYP}
In this section, we derive the four exact scar eigenstates of the PY$_4$P model [Eq. (\ref{mps_y})]. Before going to the proof, we introduce the block picture \cite{Lin2019,Shiraishi2019}. In this picture, we denote a state of the $b$-th block composed of the $(2b-1)$-th and $2b$-th spins by $\ket{O}_b=\ket{\downarrow\downarrow}_{2b-1,2b}$, $\ket{L}_b=\ket{\uparrow\downarrow}_{2b-1,2b}$, and $\ket{R}_b=\ket{\downarrow\uparrow}_{2b-1,2b}$. We note that a block $\ket{\uparrow\uparrow}_{2b-1,2b}$, which includes neighboring up spins, is prohibited. We can rewrite the four matrix product states $\ket{\Gamma_{\alpha\beta}^y}$ ($\alpha,\beta=1,2$) by
\begin{equation}
\ket{\Gamma_{\alpha\beta}^y} = \sum_{\vec{s}} \vec{v}_{\alpha}^\dagger A^{s_1} \hdots A^{s_{L_b}} \vec{v}_{\beta} \ket{\vec{s}},
\end{equation}
\begin{equation}
\begin{aligned}
A^L = \left( \begin{array}{cc}
0 & 0 \\
0 & - \sqrt{2} \end{array} \right), \quad
A^O = \left( \begin{array}{cc}
0 & -1 \\
1 & 0  \end{array} \right), \quad
A^R = \left( \begin{array}{cc}
\sqrt{2} & 0 \\
0 & 0  \end{array} \right), \quad
\end{aligned}
\vec{v}_1 = \frac{1}{\sqrt{2}} \left( \begin{array}{c}
1 \\
i \end{array} \right), \qquad
\vec{v}_2 = \frac{1}{\sqrt{2}}  \left( \begin{array}{c}
1 \\
-i \end{array} \right),
\end{equation}
where $\vec{s}=(s_1,\hdots,s_{L_b})$ represents a block configuration with $s_i=O,L,R$ and $L_b=L/2$. The relation $A^R A^L =O$ ensures the absence of adjacent up spins, Here, we show that these four states are eigenstates of the PY$_4$P Hamiltonian
\begin{eqnarray}
H_Y &=& \sum_{i=2}^{L-1} c_i  P_{i-1} Y_i P_{i+1} + Y_1 P_2 + P_{L-1} Y_L , \qquad
c_i = \sqrt{2} \cos \left( \frac{i\pi}{2} - \frac{\pi}{4} \right),
\end{eqnarray}
with eigenvalues $\{ (-1)^{\alpha-1} - (-1)^{\beta-1} \} /\sqrt{2}$. We note that these four eigenstates can be derived also from the unitary equivalence to the PXP model [Eq. (\ref{UnitaryEquiv})], but we use the block picture here so that we can treat with the PZ$_4$P model in a similar way.

\textit{Proof.---}
We consider the PY$_4$P Hamiltonian $H_Y$ based on the block picture. Let us focus on the term $h_{b,b+1}^y$, which is nontrivially acting on the $b$-th and $(b+1)$-th blocks. Then, for $b$ in the bulk ($2\leq b \leq L_b-1$),
\begin{eqnarray}
h_{b,b+1}^y &=& (-1)^b (P_{2b-1} Y_{2b} P_{2b+1} + P_{2b} Y_{2b+1} P_{2b+2}) \nonumber \\
&=&  i (-1)^b (\ket{O}\bra{R}-\ket{R}\bra{O})_b (I-\ket{L}\bra{L})_{b+1}  + i (-1)^b (I-\ket{R}\bra{R})_b (\ket{O}\bra{L}-\ket{L}\bra{O})_{b+1} \nonumber \\
&\equiv& h_{b,b+1}^{y,(2)} +  h_{b,b+1}^{y,(1)}, \\
h_{b,b+1}^{y,(2)} &=& i (-1)^b \{\ket{RL}(\bra{OL}+\bra{RO}) - \text{h.c.} \}_{b,b+1}, \label{twobody_y} \\
h_{b,b+1}^{y,(1)} &=&  i (-1)^b (\ket{O}\bra{R}-\ket{R}\bra{O})_b  + i (-1)^b (\ket{O}\bra{L}-\ket{L}\bra{O})_{b+1} \label{onebody_y}
\end{eqnarray}
are obtained. The boundary terms of $H_Y$ are given by
\begin{eqnarray}
Y_1 P_2 &=&  i (\ket{O}\bra{L}-\ket{L}\bra{O})_{1}, \qquad
P_{L-1} Y_L = - i (\ket{R}\bra{O}-\ket{O}\bra{R})_{L_b}.
\end{eqnarray}
Using the properties $A^R A^L =O$ and $A^O A^L + A^R A^O = O$ results in
\begin{equation}\label{twobody_zero_y}
h_{b,b+1}^{y,(2)}  \ket{\Gamma_{\alpha\beta}^y} = 0, \quad b=2,3,\hdots L_b-1.
\end{equation}
Thus, only the single-body terms have a nonzero contribution:
\begin{eqnarray}\label{LIOM_y}
H_Y \ket{\Gamma_{\alpha\beta}^y} = \sum_{b=1}^{L_b} h_b^{y,(1)} \ket{\Gamma_{\alpha\beta}^y}, \qquad
h_b^{y,(1)} = i (-1)^b (\ket{O}\bra{R}+\ket{L}\bra{O}-\text{h.c.})_{b}.
\end{eqnarray}
To calculate this, we consider superblocks corresponding to pairs of neighboring blocks. Here, we define a $u$-th superblock by a pair of $(2u-1)$-th and $2u$-th blocks. Each superblock has seven degrees of freedom $t_u = (s_{(2u-1)},s_{2u})=OO,LL,RR,OL,OR,LO,RO$, while $RL$ and $LR$ are not included in $\ket{\Gamma_{\alpha\beta}^y}$ because of $A^R A^L = A^L A^R = 0$. Then, the states $\ket{\Gamma_{\alpha\beta}^y}$ are rewritten in the following form,
\begin{eqnarray}
\ket{\Gamma_{\alpha\beta}^y} &=& \sum_{\vec{t}} \vec{v}_\alpha^\dagger \tilde{A}^{t_1} \hdots \tilde{A}^{t_{L_u}} \vec{v}_\beta \ket{\vec{t}}, \quad
\vec{t} = (t_1,\hdots,t_{L_u}), \quad L_u = L_b/2 = L/4.
\end{eqnarray}
The matrices $\tilde{A}^t$ are given by
\begin{eqnarray}
\begin{aligned}
& \tilde{A}^t = A^s A^{r}, \quad t=(s,r), \\
& \tilde{A}^{OO} = -\sigma^0, \, \tilde{A}^{LL} = \sigma^0 - \sigma^z, \, \tilde{A}^{RR} = \sigma^0 + \sigma^z, \\
& \tilde{A}^{OL} = -\tilde{A}^{RO} =  (\sigma^x+i\sigma^y)/\sqrt{2}, \quad
 \tilde{A}^{OR} = -\tilde{A}^{LO} = (\sigma^x-i\sigma^y) /\sqrt{2}.
\end{aligned}
\end{eqnarray}
The action of $h_{2u-1}^{y,(1)}+h_{2u}^{y,(1)}$ is described by
\begin{equation}
(h_{2u-1}^{y,(1)}+h_{2u}^{y,(1)}) \ket{\Gamma_{\alpha\beta}^y} = \sum_{\vec{t}} \vec{v}_\alpha^\dagger \tilde{A}^{t_1} \hdots \tilde{F}^{t_u} \hdots \tilde{A}^{t_{L_u}} \vec{v}_\beta \ket{\vec{t}},
\end{equation}
\begin{equation}
\tilde{F}^{(s,r)} = \left[\sum_{s^\prime} (h_{2u-1}^{y,(1)})_{ss^\prime} A^{s^\prime} \right] A^r
+ A^s \left[\sum_{s^\prime} (h_{2u}^{y,(1)})_{rr^\prime} A^{r^\prime} \right],
\end{equation}
\begin{eqnarray}
\begin{aligned}
& \tilde{F}^{OO} = O, \quad \tilde{F}^{LL} = - \tilde{F}^{RR} = - \sqrt{2} i \sigma^x, \quad
\tilde{F}^{OL} = \tilde{F}^{OR} = - \tilde{F}^{LO} = -\tilde{F}^{RO} = - i \sigma^z.
\end{aligned}
\end{eqnarray}
When we define $Y=\sigma^y/\sqrt{2}$, the matrices $\tilde{A}^{t}$ and $\tilde{F}^{t}$ are related as follows:
\begin{equation}
\tilde{F}^t  = Y \tilde{A}^t - \tilde{A}^t Y, \quad t=OO,LL,RR,OL,OR,LO,RO.
\end{equation}
Finally, we obtain
\begin{eqnarray}
H_Y \ket{\Gamma_{\alpha\beta}^y} &=& \sum_{b=1}^{L_b} h_b^{y,(1)} \ket{\Gamma_{\alpha\beta}^y} \nonumber \\
&=& \sum_{u, \vec{t}} \vec{v}_\alpha^\dagger \tilde{A}^{t_1} \hdots (Y\tilde{A}^{t_u}-\tilde{A}^{t_u}Y) \hdots \tilde{A}^{t_{L_u}} \vec{v}_\beta \ket{\vec{t}} \nonumber \\
&=& \sum_{\vec{t}} \vec{v}_\alpha^\dagger Y \tilde{A}^{t_1} \hdots \tilde{A}^{t_{L_u}} \vec{v}_\beta \ket{\vec{t}}  -  \sum_{\vec{t}} \vec{v}_\alpha^\dagger \tilde{A}^{t_1} \hdots \tilde{A}^{t_{L_u}} Y \vec{v}_\beta \ket{\vec{t}} \label{YA_AY_Appendix} \\
&=& \frac{1}{\sqrt{2}} \{ (-1)^{\alpha-1} - (-1)^{\beta-1} \} \ket{\Gamma_{\alpha\beta}^y}.
\end{eqnarray}
In the last equality, we have used the fact that $\vec{v}_{1,2}$ are eigenvectors of $Y$.  We note that the obtained formula Eq. (\ref{YA_AY_Appendix}) is equivalent to Eq. (\ref{YA_AY}) in the main text. As a result, we obtain the four exact scar eigenstates of the PY$_4$P model. $\qquad \square$

\subsection{Local conserved quantities within the embedded subspace}\label{Appendix_LocalConserved}
We would like to remark on the existence of local conserved quantities within the embedded subspace $\mathcal{S}$, spanned by $\ket{\Gamma_{\alpha\beta}^y}$. In the middle of the proof for the PY$_4$P model, we have obtained Eq. (\ref{twobody_zero_y}), which implies that the two body terms disappear when $H_Y$ acts on a state in $\mathcal{S}$. In other words, a set of local quantities $\{ h_b^{y,(1)} \}_{b=1}^{L_b}$, given by Eq. (\ref{LIOM_y}), are conserved within the subspace $\mathcal{S}$, and this is why thermalization is absent in $\mathcal{S}$ under the PY$_4$P Hamiltonian. In a similar way, the PXP model also possesses a set of local conserved quantities
\begin{equation}
h_b^{x,(1)} = (\ket{O}\bra{L}+\ket{O}\bra{R}+\text{h.c.})_b.
\end{equation}
 if we limit the Hilber space to the same subspace $\mathcal{S}$ \cite{Lin2019,Shiraishi2019}. Let us consider our periodically driven model, composed of the PXP model and the PY$_4$P model [Eq. (\ref{BinaryXY})]. Then, within the embedded subspace $\mathcal{S}$, the Floquet operator $U_f$ [Eq. (\ref{FloquetOpXY})] becomes equivalent to
 \begin{equation}
U_f |_\mathcal{S} = \prod_{b=1}^{L_b} \left( e^{-i T_2 h_b^{y,(1)}} e^{-i T_1 h_b^{x,(1)}} \right).
\end{equation}
This indicates that, focusing on the stroboscopic dynamics,  the model possesses a macroscopic number of local conserved quantities $\{ -i \log ( \exp(-i T_2 h_b^{y,(1)}) \exp(-i T_1 h_b^{x,(1)}) ) \}_b$ only within $\mathcal{S}$, and hence any state in $\mathcal{S}$ does not experience thermalization to infinite temperature.
 
 We also note that this property makes the nonlocal chiral symmetry operator $\mathcal{C}$ [Eq. (\ref{ChiralSymmetry})] local within the embedded subspace $\mathcal{S}$:
\begin{equation}
\mathcal{C} = \left( \prod_i Z_i \right) \exp \left( i \sum_b h_b^{y,(1)} T_2 \right).
\end{equation}
As discussed in Section \ref{Sec_ExactScar} in the main text, this indicates that the two exact Floquet scar eigenstates which are related to each other by $\mathcal{C}$ possess the same entanglement entropy, while other pairs outside of $\mathcal{S}$ do not due to the nonlocality of $\mathcal{C}$.

\subsection{Exact scar eigenstates of the PZ$_4$P model}\label{Appendix_PZP}
In this section, we show that the four matrix product states $\ket{\Gamma_{\alpha\beta}^z}$ ($\alpha,\beta=1,2$), given by Eq. (\ref{mps_z}), are eigenstates of the PZ$_4$P Hamiltonian
\begin{equation}
H_Z = -\sqrt{2} \left( \sum_{i=2}^{L-1} c_i P_{i-1} Q_i P_{i+1} + Q_1 P_2 + P_{L-1} Q_L \right), \quad
Q_i = I_i - P_i = (1+Z_i)/2.
\end{equation}
Using the block picture introduced in Appendix \ref{Appendix_PYP}, $\ket{\Gamma_{\alpha\beta}^z}$ is rewritten as follows:
\begin{equation}
\ket{\Gamma_{\alpha\beta}^z} = \sum_{\vec{s}} \vec{w}_{\alpha}^\dagger A^{s_1} \hdots A^{s_{L_b}} \vec{w}_{\beta} \ket{\vec{s}}, \quad \vec{w}_1 = \left( \begin{array}{c}
1 \\
0 \end{array} \right), \quad
\vec{w}_2 = \left( \begin{array}{c}
0 \\
1 \end{array} \right).
\end{equation}
We derive that these four states possess eigenvalues $\{ (-1)^{\alpha-1} - (-1)^{\beta-1} \} /\sqrt{2}$ below.

\textit{Proof.---}The proof goes in a similar way to the one for the PY$_4$P model. The terms acting on the $b$-th and $(b+1)$-th blocks in $H_Z$, denoted by $h_{b,b+1}$, are given by
\begin{eqnarray}
h_{b,b+1}^z &=& \sqrt{2} (-1)^{b-1} (P_{2b-1} Q_{2b} P_{2b+1} + P_{2b} Q_{2b+1} P_{2b+2}) \nonumber \\
&=& \sqrt{2} (-1)^{b-1} (\ket{R}\bra{R})_b (I-\ket{L}\bra{L})_{b+1} + \sqrt{2} (-1)^{b-1} (I-\ket{R}\bra{R})_b (\ket{L}\bra{L})_{b+1} \nonumber \\
&\equiv& h_{b,b+1}^{z,(2)} +  h_{b,b+1}^{z,(1)}, \\
h_{b,b+1}^{z,(2)} &=& -2 \sqrt{2} (-1)^{b-1} (\ket{RL}\bra{RL})_{b,b+1}, \quad
h_{b,b+1}^{z,(1)} = \sqrt{2} (-1)^{b-1} (\ket{R}\bra{R})_b + \sqrt{2} (-1)^{b-1} (\ket{L}\bra{L})_{b+1}, \label{onebody_z}
\end{eqnarray}
and the boundary terms are
\begin{equation}
Q_1 P_2 = -\sqrt{2} (\ket{L}\bra{L})_1, \quad P_{L-1} Q_L = - \sqrt{2} (\ket{R}\bra{R})_{L_b}.
\end{equation}
The property $A^R A^L = O$ results in $h_{b,b+1}^{z,(2)} \ket{\Gamma_{\alpha\beta}^z} =0$, and hence
\begin{eqnarray}
H_Z \ket{\Gamma_{\alpha\beta}^z} &=& \sum_{b=1}^{L_b} h_b^{z,(1)} \ket{\Gamma_{\alpha\beta}^z},  \quad
h_b^{z,(1)} = \sqrt{2} (-1)^b (\ket{L}\bra{L}-\ket{R}\bra{R})_{b}
\end{eqnarray}
is obtained. When we move to the superblock picture, the action of $h_{2u-1}^{z,(1)}+h_{2u}^{z,(1)}$ can be calculated:
\begin{equation}
(h_{2u-1}^{z,(1)}+h_{2u}^{z,(1)}) \ket{\Gamma_{\alpha\beta}^z} = \sum_{\vec{t}} \vec{w}_\alpha^\dagger \tilde{A}^{t_1} \hdots \tilde{G}^{t_u} \hdots \tilde{A}^{t_{L_u}} \vec{w}_\beta \ket{\vec{t}},
\end{equation}
\begin{equation}
\tilde{G}^{(s,r)} = \left[\sum_{s^\prime} (h_{2u-1}^{z,(1)})_{ss^\prime} A^{s^\prime} \right] A^r
+ A^s \left[\sum_{s^\prime} (h_{2u}^{z,(1)})_{rr^\prime} A^{r^\prime} \right].
\end{equation}
Calculating $\tilde{G}^{t}$ respectively for $t=(s,r)$ results in
\begin{equation}
\tilde{G}^{OO}=\tilde{G}^{LL}=\tilde{G}^{RR}=O, \quad \tilde{G}^{OL}=-\tilde{G}^{LO}= \sigma^x + i \sigma^y, \quad \tilde{G}^{OR}=-\tilde{G}^{RO}=-(\sigma^x-i\sigma^y).
\end{equation}
One can confirm the relation between $\tilde{G}^{t}$ and $\tilde{A}^{t}$ as
\begin{equation}
\tilde{G}^t  = Z \tilde{A}^t - \tilde{A}^t Z, \quad t=OO,LL,RR,OL,OR,LO,RO,
\end{equation}
where $Z$ is defined by $Z=\sigma^z/\sqrt{2}$. Therefore, we obtain
\begin{eqnarray}
H_Z \ket{\Gamma_{\alpha\beta}^z} 
&=& \sum_{\vec{t}} \vec{w}_\alpha^\dagger Z \tilde{A}^{t_1} \hdots \tilde{A}^{t_{L_u}} \vec{w}_\beta \ket{\vec{t}}  -  \sum_{\vec{t}} \vec{w}_\alpha^\dagger \tilde{A}^{t_1} \hdots \tilde{A}^{t_{L_u}} Z \vec{w}_\beta \ket{\vec{t}} \label{ZA_AZ} \\
&=& \frac{1}{\sqrt{2}} \{ (-1)^{\alpha-1} - (-1)^{\beta-1} \} \ket{\Gamma_{\alpha\beta}^z},
\end{eqnarray}
and hence $ \ket{\Gamma_{\alpha\beta}^z}$ are eigenstates of $H_Z$. $\qquad \square$

\subsection{Relationship among the PXP model, PY$_4$P model and PZ$_4$P model}
We would like to note the relationship among the static models and explain how the PY$_4$P model and the PZ$_4$P model are constructed. First, let us introduce an embedded Hamiltonian, with which exact scar eigenstates can be systematically obtained \cite{Shiraishi2017}. An embedded Hamiltonian with an embedded subspace $\mathcal{T}$ is given by
\begin{equation}\label{EmbeddedHamiltonian}
H_\text{em}= \sum_i P_i h_i P_i + H^\prime, \quad
P_i \mathcal{T} = 0 \, \, (^\forall i), \quad H^\prime \mathcal{T} \subset \mathcal{T},
\end{equation}
where $P_i$ is a projection operator. Although the nonintegrability is still nontrivial, the subspace $\mathcal{T}$ is immune to thermalization since the dynamics under $H_\text{em}$ is closed within it. It is known that the PXP Hamiltonian $H_X$ can be transformed to a certain embedded Hamiltonian with the embedded subspace $\mathcal{S}=\mathrm{span} \{ \ket{\Gamma_{11}^x}, \ket{\Gamma_{12}^x}, \ket{\Gamma_{21}^x}, \ket{\Gamma_{22}^x} \}$ \cite{Shiraishi2019}. There are options of Hermitian operators $h_i$ and $H^\prime$ in Eq. (\ref{EmbeddedHamiltonian}), as long as $P_i \mathcal{T}=0$ and $H^\prime \mathcal{T} \subset \mathcal{T}$ are satisfied. Ref. \cite{Shiraishi2019} claims that these options enable one to construct generalized versions of the PXP Hamiltonian, which show exact quantum many-body scars. The PY$_4$P Hamiltonian $H_Y$, the PZ$_4$P Hamiltonian $H_Z$, and their linear combinations $\vec{a} \cdot \vec{H}$ exemplify the generalized versions, and we can obtain them by imposing a Rydberg blockade, which prohibits generation of adjacent up spins (i.e. PXP-type Hamiltonians), on the option of $h_i$ and $H^\prime$.

\subsection{Observables}\label{AppendixEmbedded}
In this section, we calculate matrix elements of observables within the embedded subspace $\mathcal{S}$ spanned by $\{ \ket{\Gamma_{\alpha\beta}^\nu} \}_{\alpha,\beta=1,2}$. We consider a certain observable $O_b$ acting on the $b$-th block, and then we define
\begin{equation}
F^{s} = \sum_{s^\prime=O,L,R} (O_b)_{s s^\prime} A^{s^\prime}.
\end{equation}
We can calculate the matrix elements $\braket{\Gamma_{\alpha\beta}^\nu | O_b | \Gamma_{\alpha^\prime \beta^\prime}^\nu}$ by
\begin{eqnarray}
\braket{\Gamma_{\alpha\beta}^\nu | O_b | \Gamma_{\alpha^\prime \beta^\prime}^\nu}  &=& 
\sum_{\vec{s}} \{ (\vec{u}_\alpha^\nu)^\dagger A^{s_1} \hdots A^{s_{L_b}} \vec{u}_\beta^\nu \}^\ast 
\{ (\vec{u}_{\alpha^\prime}^\nu)^\dagger A^{s_1} \hdots F^{s_b} \hdots A^{s_{L_b}} \vec{u}_{\beta^\prime}^\nu \} \nonumber \\
&=& (\vec{U}_{\alpha\alpha^\prime}^\nu )^\dagger (E_{AA})^b E_{AF} (E_{AA})^{L_b-b-1} (\vec{U}_{\beta\beta^\prime}^\nu ), \\
E_{AA} &\equiv& \sum_{s} (A^{s})^\ast \otimes A^{s} = \left( \begin{array}{cccc}
2 & 0 & 0 & 1 \\
0 & 0 & -1 & 0 \\
0 & -1 & 0 & 0 \\
1 & 0 & 0 & 2
\end{array} \right),
\quad
E_{AF} \equiv \sum_{s} (A^{s})^\ast \otimes F^{s}, \quad
\vec{U}_{\alpha\alpha^\prime}^\nu \equiv (\vec{u}_\alpha^\nu)^\ast \otimes \vec{u}_{\alpha^\prime}^\nu.
\end{eqnarray}

The norm and the overlap of $\{ \ket{\Gamma_{\alpha\beta}^\nu} \}_{\alpha,\beta=1,2}$ are evaluated by setting $O_b=I_b$, which results in
\begin{equation}
\braket{\Gamma_{11}^\nu | \Gamma_{11}^\nu} = \braket{\Gamma_{22}^\nu | \Gamma_{22}^\nu} = \frac{1}{2} (3^{L_b} +1), \quad 
\braket{\Gamma_{12}^\nu | \Gamma_{12}^\nu} = \braket{\Gamma_{21}^\nu | \Gamma_{21}^\nu} = \frac{1}{2} (3^{L_b} -1),
\end{equation}
\begin{equation}
\braket{\Gamma_{11}^\nu | \Gamma_{12}^\nu} = \braket{\Gamma_{11}^\nu | \Gamma_{21}^\nu} =
\braket{\Gamma_{22}^\nu | \Gamma_{12}^\nu} = \braket{\Gamma_{22}^\nu | \Gamma_{21}^\nu} =
\braket{\Gamma_{12}^\nu | \Gamma_{21}^\nu} = 0, \quad
\braket{\Gamma_{11}^\nu | \Gamma_{22}^\nu} = 1
\end{equation}
 for even $L_b$ and $\nu=x,y,z$. Thus, under the renormalization by $\ket{\tilde{\Gamma}_{\alpha\beta}^\nu} \equiv \ket{\Gamma_{\alpha\beta}^\nu} / || \ket{\Gamma_{\alpha\beta}^\nu}  ||$, $\{  \ket{\tilde{\Gamma}_{\alpha\beta}^\nu} \}_{\alpha,\beta=1,2}$ is an orthonormal basis of the embedded subspace $\mathcal{S}$.
 
The Pauli $x$ [$z$] operator for the $b$-th block is represented by $O_b=(\ket{L}\bra{O}+\ket{R}\bra{O}+\text{h.c.})_b$ [ $O_b=-2(\ket{O}\bra{O})_b$ ] The matrix representation under the basis $\{  \ket{\tilde{\Gamma}_{\alpha\beta}^x} \}_{\alpha,\beta=1,2}$ in the thermodynamic limit is given by the following equation:
 \begin{equation}
\lim_{L \to \infty} \sum_{i=1}^L \braket{\tilde{\Gamma}_{\alpha\beta}^x | X_i | \tilde{\Gamma}_{\alpha^\prime \beta^\prime}^x} =  \frac{1}{\sqrt{2}} \{ (-1)^{\alpha-1} - (-1)^{\beta-1} \} \delta_{\alpha \alpha^\prime} \delta_{\beta \beta^\prime}, \quad
\lim_{L \to \infty} \frac{1}{L} \sum_{i=1}^L \braket{\tilde{\Gamma}_{\alpha\beta}^x | Z_i | \tilde{\Gamma}_{\alpha^\prime \beta^\prime}^x} =  -\frac{1}{6} \delta_{\alpha \alpha^\prime} \delta_{\beta \beta^\prime}.
\end{equation}
On the other hand, the domain-wall density $D_b=(I_b-Z_{2b-1}Z_{2b})/2$ is obtained by setting $O_b=(\ket{L}\bra{L}+\ket{R}\bra{R})_b$. The matrix representation under the basis $\{  \ket{\tilde{\Gamma}_{\alpha\beta}^\nu} \}_{\alpha,\beta=1,2}$ in the thermodynamic limit is  \begin{equation}
\lim_{L_b \to \infty} \braket{\tilde{\Gamma}_{\alpha\beta}^\nu | D_b | \tilde{\Gamma}_{\alpha^\prime \beta^\prime}^\nu} = \frac{2}{3} \delta_{\alpha \alpha^\prime} \delta_{\beta \beta^\prime}, \quad \nu=x,y,z.
\end{equation}

\subsection{Floquet intrinsic scar eigenstate}\label{FloquetIntrinsic}
We discuss the relation between our model and the \textit{Floquet-intrinsic scar}, the exact Floquet quantum many-body scar recently proposed by Sugiura \textit{et al}. \cite{Sugiura2019}. They regard Floquet scar eigenstates which become simultaneous eigenstates of all the instantaneous frustration-free Hamiltonians $H(t)$ as trivial ones. In contrast to this, in the case of a binary drive Eq. (\ref{BinaryDrive}), a Floquet-intrinsic scar eigenstate is defined by a scar eigenstate which is neither an eigenstate of $H_1$ nor that of $H_2$, but a simultaneous eigenstate of their time evolution operators $\exp(-iH_1T_1)$ and $\exp(-iH_2T_2)$. Coexistence of these conditions is brought by the equivalence of quasienergy modulo $2\pi$ in Floquet systems, and hence Floquet-intrinsic scar eigenstates are unique to Floquet systems.

In our model, the Floquet eigenstate $\ket{\Gamma_0}$ is a simultaneous eigenstate of the instanteous Hamiltonians $H_X$ and $H_Y$. In general, the other three are neither an eigenstate of any instantaneous Hamiltonian nor that of its time evolution operator. Under fine-tuning of the parameters, Floquet-intrinsic scar eigenstates can appear in our model Eq. (\ref{BinaryXY}). Assume that the durations $T_1$ and $T_2$ are fixed as follows:
\begin{equation}\label{FineTune}
T_1 = \frac{(2m-1)\pi}{\sqrt{2}}, \quad T_2 = \frac{(2n-1)\pi}{\sqrt{2}}, \quad m, n \in \mathbb{N}.
\end{equation}
Then, using Eq. (\ref{eigen_yx}), we find the four simultaneous eigenstates  of $\exp(-iH_XT_1)$ and $\exp(-iH_YT_2)$. One is given by Eq. (\ref{Gamma_zero}), and the other three $\ket{\Gamma(u_X,u_Y)}$ are as follows:
\begin{eqnarray}
\begin{aligned}
& \ket{\Gamma(+1,-1)} = \ket{\Gamma_{11}^x} - \ket{\Gamma_{22}^x} = i( \ket{\Gamma_{12}^y} - \ket{\Gamma_{21}^y}), \\
& \ket{\Gamma(-1,+1)} =\ket{\Gamma_{12}^x} - \ket{\Gamma_{21}^x} = i (\ket{\Gamma_{11}^y} - \ket{\Gamma_{22}^y}), \\
& \ket{\Gamma(-1,-1)} = \ket{\Gamma_{12}^x} + \ket{\Gamma_{21}^x} = \ket{\Gamma_{12}^y} + \ket{\Gamma_{21}^y},
\end{aligned}
\end{eqnarray}
where the indices $u_X, u_Y = \pm 1$ represent eigenvalues of $\exp(-iH_XT_1)$ and $\exp(-iH_YT_2)$ respectively. The state $\ket{\Gamma(-1,-1)}$, corresponding to the eigenstate Eq. (\ref{tilde_Gamma_zero}), is neither an eigenstate of $H_X$ nor that of $H_Y$, and it is a Floquet-intrinsic scar eigenstate. This Floquet-intrinsic scar eigenstate appears due to the emergent degeneracy of $\ket{\Gamma_{12}^x}$ and $\ket{\Gamma_{21}^x}$ under $\exp(-iH_XT_1)$ and that of $\ket{\Gamma_{12}^y}$ and $\ket{\Gamma_{21}^y}$ under $\exp (-iH_YT_2)$, although they are not originally degenerated under $H_X$ or $H_Y$. This degeneracy is caused by equivalence of quasienergy modulo $2\pi$ unique to Floquet systems.

Floquet-intrinsic scar eigenstates are fragile to the change of parameters. In the originally proposed model by Sugiura \textit{et al.} \cite{Sugiura2019}, all the Floquet scar eigenstates found rigorously are Floquet-intrinsic scar eigenstates, and hence quantum many-body scars are not observed in the absence of fine tuning. On the other hand, in our case, breakdown of the condition Eq. (\ref{FineTune}) causes disappearance of the Floquet-intrinsic eigenstate as well, but the Floquet many-body scars still exist. In other words, our model includes Floquet-intrinsic scars as a special choice of the parameters.

\section{Properties of Infinite temperature states}
For this paper to be self-contained, we summarize the properties of the infinite temperature state in the constrained Hilbert space $\mathcal{H}$ under Rydberg blockade \cite{Turner2018B}. We denote its dimension for the system size $L$ under open boundary conditions by $\mathcal{D}_L$. The condition that pairs of adjacent up spins are prohibited results in
\begin{equation}
\mathcal{D}_1 = 2, \quad \mathcal{D}_2 = 3, \quad \mathcal{D}_{L+2} = \mathcal{D}_{L+1} + \mathcal{D}_{L}.
\end{equation}
This is nothing but the definition of the Fibonacci sequence, and hence we obtain
\begin{equation}
\mathcal{D}_L = \frac{1}{\sqrt{5}} \left\{ \phi^{L+2} - (1-\phi)^{L+2} \right\}, \quad \phi = \frac{1+\sqrt{5}}{2}.
\end{equation}

\subsection{Entanglement entropy}
We evaluate the entanglement entropy at infinite temperature when we split the system in half. Denoting the left half (the right half) by $A$ ($B$), the reduced density operator of the infinite temperature state $\rho_\infty=I_{\mathcal{D}_L}/\mathcal{D}_L$ for the subsystem $A$ is
\begin{equation}
\rho_\infty^A \equiv \mathrm{Tr}_B [\rho_\infty] = \frac{1}{\mathcal{D}_L} \left( \mathcal{D}_{L/2-1} \sum_{\vec{\sigma} \in \mathcal{K}_\uparrow} \ket{\vec{\sigma}} \bra{\vec{\sigma}} + \mathcal{D}_{L/2} \sum_{\vec{\sigma} \in \mathcal{K}_\downarrow} \ket{\vec{\sigma}} \bra{\vec{\sigma}} \right).
\end{equation}
Here, we define $\mathcal{K}_\uparrow$ ($\mathcal{K}_\downarrow$) by a set of configurations of $L/2$ spins, the spin of which at the right edge is $\uparrow$ ($\downarrow$).
Using the equations $|\mathcal{K}_\uparrow|=\mathcal{D}_{L/2-2}$ and $|\mathcal{K}_\downarrow|=\mathcal{D}_{L/2-1}$, we obtain the entanglement entropy at infinite temperature as follows:
\begin{equation}
S_\infty \equiv - \mathrm{Tr}_A [ \rho_\infty^A \log \rho_\infty^A ] = - \frac{\mathcal{D}_{L/2-2}\mathcal{D}_{L/2-1}}{\mathcal{D}_L} \log \left( \frac{\mathcal{D}_{L/2-1}}{\mathcal{D}_L} \right)
- \frac{\mathcal{D}_{L/2-2}\mathcal{D}_{L/2-1}}{\mathcal{D}_L} \log \left( \frac{\mathcal{D}_{L/2-1}}{\mathcal{D}_L} \right),
\end{equation}
and in the thermodynamic limit, the entanglement entropy per volume becomes
\begin{equation}
\lim_{L\to\infty}\frac{S_\infty}{L} =\frac{1}{2} \log \phi.
\end{equation}

\subsection{Observables}\label{AppendixInfinite}
The expectation value of a certain observable $O$ at infinite temperature is given by
\begin{eqnarray}
\left< O \right>_{T=\infty} &\equiv& \frac{1}{\mathcal{D}_L} \mathrm{Tr}_\mathcal{H} [O]
= \frac{1}{\mathcal{D}_L} \sum_{\vec{\sigma} \in \mathcal{K}_L} \braket{\vec{\sigma} | O | \vec{\sigma}},
\end{eqnarray}
where $\mathcal{K}_L$ represents a set of classical spin configurations of an $L$-site chain which includes no adjacent up spins.

We here discuss the expectation values of the Pauli operators $X_i$ and $Z_i$, and the domain-wall density $D_b$. Since the operator $X_i$ has only off-diagonal elements in the basis $\{ \ket{\vec{\sigma}} \}_{\vec{\sigma}\in \mathcal{K}_L}$, we obtain
\begin{eqnarray}
\left< X_i \right>_{T=\infty} &=& \frac{1}{\mathcal{D}_L} \sum_{\vec{\sigma} \in \mathcal{K}_L} \braket{\vec{\sigma} | X_i | \vec{\sigma}} =0.
\end{eqnarray}

Next, we consider the Pauli $z$ operator $Z_i$. When we fix the $i$-th spin by $\uparrow$ ($\downarrow$), the number of possible spin configurations is $\mathcal{D}_{i-2} \times \mathcal{D}_{L-i-1}$ ($\mathcal{D}_{i-1} \times \mathcal{D}_{L-i}$). Therefore, we obtain the expectation value for finite-size and infinite-size systems as follows:
\begin{equation}
\left< Z_i \right>_{T=\infty} = \frac{1}{\mathcal{D}_L} \sum_{\vec{\sigma} \in \mathcal{K}_L} \braket{\vec{\sigma} | Z_i | \vec{\sigma}} 
=  \frac{\mathcal{D}_{i-2} \mathcal{D}_{L-i-1} - \mathcal{D}_{i-1} \mathcal{D}_{L-i}}{\mathcal{D}_L}, \quad
\lim_{L\to \infty} \left< Z_{L/2} \right>_{T=\infty} = - \frac{1}{\sqrt{5}}.
\end{equation}

In a similar way, we obtain the domain-wall density as follows:
\begin{equation}
\left< D_b \right>_{T=\infty} = \frac{1}{\mathcal{D}_L} \sum_{\vec{\sigma} \in \mathcal{K}_L} \braket{\vec{\sigma} | D_b | \vec{\sigma}}
= \frac{\mathcal{D}_{2b-3} \mathcal{D}_{2L_b-2b} + \mathcal{D}_{2b-2} \mathcal{D}_{2L_b-2b-1}}{\mathcal{D}_{2L_b}}, \quad
\lim_{L_b \to \infty} \left< D_{L_b/2} \right>_{T=\infty} = \frac{2}{\sqrt{5}\phi}.
\end{equation}

\section{Dynamics of Special scar states}\label{SpecialScar}
\begin{figure*}
\hspace{-1cm}
\begin{center}
    \includegraphics[height=6.5cm, width=18cm]{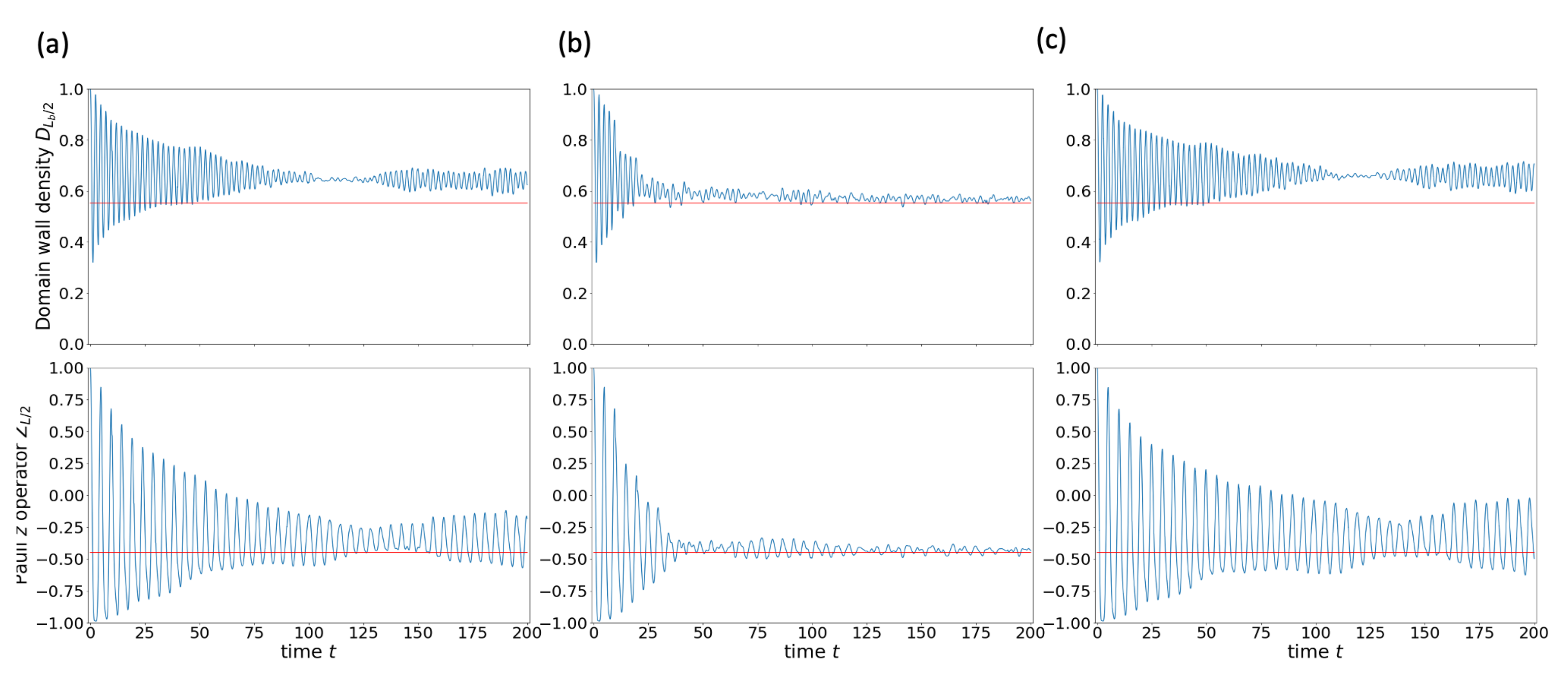}
    \caption{Real-time dynamics under the special initial state $\ket{\mathbb{Z}_2}=\ket{\uparrow\downarrow\uparrow\downarrow\hdots}$: (a) under the static PY$_4$P Hamiltonian, or equivalently at $T_1=10, T_2=0$l (b) under the binary drive at $T_1=9.5, T_2=0.5$; and (c) under the binary drive at $T_1=0.95, T_2=0.05$. (a) Both the domain-wall density and the Pauli $z$ operator show long-lasting oscillations without approaching their thermal equilibrium values at $T=\infty$. (b) The observables rapidly approach those of the infinite temperature due to the drive. (c) Nonthermalizing behaviors of the observables are observed as in the static case in spite of the existence of the drive. These are brought by pre-equilibration under an effective static Hamiltonian in the high-frequency regime of Floquet systems.} 
    \label{Fig_special_dynamics}
 \end{center}
 \end{figure*}

In this section, we discuss the relationship between the model showing exact Floquet many-body scars and a nonthermalizing oscillation of observables in the static PXP model \cite{Bernien2017,Turner2018}.
In the PXP model, the embedded subspace spanned by the exact scar eigenstates is perfectly immune to thermalization. However, there also exist some special scar states seemingly immune to thermalization although they are not included in the embedded subspace. In fact, an extremely long-term oscillation of the domain-wall density is observed under the preparation of special initial states such as $\ket{\mathbb{Z}_2}=\ket{\uparrow\downarrow\uparrow\downarrow\hdots}$ and $\ket{\mathbb{Z}_3}=\ket{\uparrow\downarrow\downarrow\uparrow\downarrow\downarrow\hdots}$ in Rydberg atoms. Our periodically-driven model is composed of the PXP Hamiltonian $H_X$ and the PY$_4$P Hamiltonian $H_Y$, and each of them shows nonthermalizing oscillations of observables under the specific initial-state preparation (See Fig. \ref{Fig_special_dynamics} (a) for the nonthermal behavior of the PY$_4$P model). Here, we numerically examine whether a nonthermalizing oscillation appears also in the driven cases, and discuss its origin.

Figure \ref{Fig_special_dynamics} shows the real-time dynamics when we begin with the $\mathbb{Z}_2$ ordered state $\ket{\mathbb{Z}_2}$ [See (b) and (c)]. Figure \ref{Fig_special_dynamics} (c) indicates that the model for the relatively-small period $T=T_1+T_2=1$ shows nonthermalizing oscillations of the domain-wall density $D_{L/2}$ and the Pauli operator $Z_{L/2}$. These nonthermalizing behaviors are expected to originate from pre-equilibration of Floquet systems in the high-frequency regime \cite{Kuwahara2016,Mori2016,Abanin2017B,Abanin2017Mat}.  When the local energy scale of the Hamiltonian is small enough compared to the frequency, its stroboscopic dynamics is well described by a static effective Hamiltonian given by the Floquet-Magnus expansion. Up to the lowest order in $T$, the static effective Hamiltonian for our model is given by the time-averaged one over one period,
\begin{equation}
H_\text{eff} = \frac{T_1}{T_1+T_2} H_X +  \frac{T_2}{T_1+T_2} H_Y + O(T).
\end{equation}
Thus, through Fig.  \ref{Fig_special_dynamics} (c), we observe nonthermalizing behaviors caused by static quantum many-body scars in the periodically-driven model.

On the other hand, when the local energy scale is comparable to the frequency or larger than it, we expect that effective static behaviors do not appear. Figure  \ref{Fig_special_dynamics} (b) shows the dynamics for such a Floquet intrinsic regime, where the local energy scale $1$ is larger than the frequency $2\pi/T=\pi/5$. This result represents that the domain-wall density $D_{L/2}$ and the Pauli operator $Z_{L/2}$ quickly relax to the values of infinite temperature states, and that the nonthermalizing oscillations disappear by the periodic drive in spite of the instantaneous Hamiltonians $H_X$ and $H_Y$. We focus on the origin of this behavior below.

\begin{figure*}
\hspace{-1cm}
\begin{center}
    \includegraphics[height=4.5cm, width=18cm]{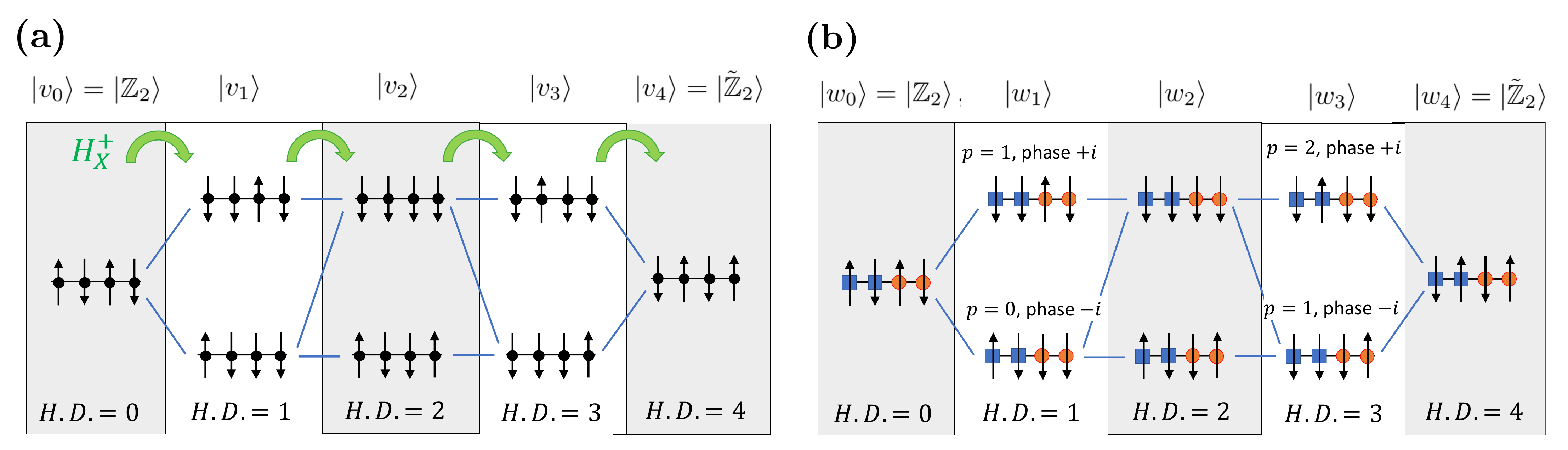}
    \caption{Schematic pictures of approximately closed subspaces obtained by FSA: (a) for the PXP model and (b) for the PY$_4$P model at $L=4$. Each layer denoted by $\ket{v_n}$ or $\ket{w_n}$ is composed of product states the Hamming distance of which from $\ket{\mathbb{Z}_2}$ is fixed to $n$. In the PY$_4$P model (b), blue squares and orange circles represent additional phases $+i$ and $-i$ obtained when the spins at their positions are flipped. The term ``phase''  at each state means its coefficient due to these additional phases.  } 
    \label{Fig_FSA}
 \end{center}
 \end{figure*}

We expect that this can be explained in terms of forward scattering approximation (FSA) \cite{Turner2018,Turner2018B}. By means of FSA, we can obtain an approximately closed subspace of the dynamics under a certain initial state, and write down the effective Hamiltonian within this subspace. In the case of the PXP model,  the Hamiltonian $H_X$ can be divided into two terms as follows:
\begin{eqnarray}
H_X &=& H_X^+ + H_X^-, \quad H_X^+ = \sum_{i:\text{odd}} P_{i-1} S_i^- P_{i+1} + \sum_{i:\text{even}} P_{i-1} S_i^+ P_{i+1}, \quad H_X^- = (H_X^+)^\dagger, \\
S_i^\pm &=& \frac{1}{2}(X_i \pm i Y_i), \quad P_0=P_{L+1}=I
\end{eqnarray}
under open boundary conditions. When we begin with $\ket{\mathbb{Z}_2}$, we define a set of orthonormal states by
\begin{equation}
\ket{v_0}=\ket{\mathbb{Z}_2}, \quad \ket{v_n} = \frac{(H_X^+)^n \ket{v_0}}{|| (H_X^+)^n \ket{v_0} ||}, \quad n=0,1,2,\hdots, L.
\end{equation}
When we denote the Hamming distance (the smallest number of spin flips required to convert two states) from $\ket{\mathbb{Z}_2}$ by $H.D.$, the term $H_X^+$ ($H_X^-$) increases (decreases) $H.D.$ by one. Thus, the state $\ket{v_n}$ becomes a superposition of states the spin configurations of which satisfy $H.D.=n$, and thereby we obtain $\ket{v_L}=\ket{\tilde{\mathbb{Z}}_2}$ and $\ket{v_{L+1}}$=0. From the numerical calculation up to $L=32$ \cite{Turner2018B}, it is known that the dynamics from the initial state $\ket{\mathbb{Z}_2}$ under $H_X$ is approximately closed within the subspace $\mathcal{R}_X$, spanned by $\left\{\ket{v_n} \right\}_{n=0}^L$ [See Fig. \ref{Fig_FSA} (a)]. This is one of the possible explanations for long-term nonthermalizing oscillations in the PXP model \cite{Turner2018,Turner2018B}.

In a similar way, we apply FSA to the PY$_4$P model. The Hamiltonian is written as
\begin{eqnarray}\label{HyFSA}
H_Y &=& H_Y^+ + H_Y^-, \quad H_Y^+ = i \sum_{i:\text{odd}} c_i P_{i-1} S_i^- P_{i+1} - i \sum_{i:\text{even}} c_i P_{i-1} S_i^+ P_{i+1}, \quad H_Y^- = (H_Y^+)^\dagger,
\end{eqnarray}
and then the dynamics under $H_Y$ is approximately closed within the subspace $\mathcal{R}_Y$ spanned by $\left\{ \ket{w_n} = (H_Y^+)^n \ket{\mathbb{Z}_2}/ || (H_Y^+)^n \ket{\mathbb{Z}_2} || \right\}_{n=0}^L$. A state $\ket{w_n}$ is a superposition of states the Hamming distance of which from $\ket{\mathbb{Z}_2}$ is $n$ as well. However, in the case of the PY$_4$P model, a spin flip gives additional phases $+i$ or $-i$ to each state depending on the flipped-spin's site due to the signs of $c_i$ and the coefficients in Eq. (\ref{HyFSA}) [See Fig. \ref{Fig_FSA} (b)].
Then, we can understand the thermalization to infinite temperature under the Floquet drive with the initial state $\ket{\mathbb{Z}_2}$ from the difference between the closed subspaces $\mathcal{R}_X$ and $\mathcal{R}_Y$. We immediately obtain $\ket{v_0}=\ket{w_0},\ket{v_L}=\ket{w_L}$, and $\braket{v_i | w_j} = 0$ for different $i,j$ by their definitions. First, let us consider the overlap between $\ket{v_1}$ and $\ket{w_1}$. These states are equally-weighted superpositions of states where one of the odd sites is flipped from $\ket{\mathbb{Z}_2}$. Since the number of the states with an additional phase $+i$ is equal to that of the states with an additional phase $-i$, we obtain
\begin{equation}
\braket{v_1|w_1}= \frac{1}{N_{H.D.=1}} \{ (-i) \times L/4 + i \times L/4 \} =0,
\end{equation}
where $N_{H.D.=n}$ represents the number of possible spin configurations with $H.D.=n$ under the constrained Hilbert space. Next, we consider the overlap $\braket{v_n|w_n}$ for generic $n$ in the limit of $L\to \infty$. Here, let $p$ denote the number of flipped spins with an additional phase $+i$ [the number of the blue squares flipped in Fig. \ref{Fig_FSA} (b)]. Then, each state in $\ket{w_n}$ has an additional phase given by $(+i)^p (-i)^{n-p}=(-1)^p (-i)^n$, and the overlap $\braket{v_n|w_n}$ is determined by its summation over states with $H.D.=n$. Let us consider the case  where $n$ is odd. The additional phase $(-1)^p (-i)^n$ depends on the parity of $p$. Choosing odd $p$ flipped sites with the blue squares is equivalent to choosing even $n-p$ flipped sites with orange circles. Assume that the system size $L$ is large enough, and then we can neglect the effect of the boundaries. With considering the symmetry of the blue squares and the orange circles  in the bulk, the total contributions with odd $p$ is equal to that for even $p$. Thus, we obtain 
\begin{equation}
\lim_{L\to\infty} \braket{v_n|w_n} = 0, \quad \text{for} \quad \text{odd} \quad n
\end{equation}
because $N_{H.D.=n}$ grows with increasing $L$. These macroscopic numbers (at least $L/2$) of orthogonality relations represent that a generic state in $\mathcal{R}_X$ (or $\mathcal{R}_Y$) flows out of $\mathcal{R}_X$ (or $\mathcal{R}_Y$)  under the time evolution by the Hamiltonian $H_Y$ (or $H_X$). Finally, we conclude that, under the periodic switching of $H_X$ and $H_Y$, the dynamics from $\ket{\mathbb{Z}_2}$ is no longer closed within the original subspaces $\mathcal{R}_X$ or $\mathcal{R}_Y$, and hence thermalization to infinite temperature is observed without showing nonthermalizing oscillations.
\end{document}